\documentclass[%
 reprint,
superscriptaddress,
showpacs,
 amsmath,amssymb,
 aps,
 prx,
]{revtex4-1}

\usepackage{graphicx}
\usepackage{dcolumn}
\usepackage{bm}
\usepackage{times}
\usepackage{xcolor}
\usepackage{hyperref}
\usepackage{afterpage}
\usepackage{natbib}
\usepackage{stmaryrd} 
\usepackage{MnSymbol} 

\makeatletter
\newcommand{\fixed@sra}{$\vrule height 2\fontdimen22\textfont2 width 0pt\shortrightarrow$}
\newcommand{\shortarrow}[1]{%
	\mathrel{\text{\rotatebox[origin=c]{\numexpr#1*45}{\fixed@sra}}}
}
\newcommand{\uprightarrow}{\shortarrow{1}}
\newcommand{\upleftarrow}{\shortarrow{3}}
\newcommand{\downleftarrow}{\shortarrow{5}}
\newcommand{\downrightarrow}{\shortarrow{7}}
\makeatother
\begin{document}


\title{Hierarchy of energy scales in an O(3) symmetric
antiferromagnetic quantum critical metal: \\ a Monte Carlo study}

\author{Carsten Bauer}
\affiliation{Institute for Theoretical Physics, University of Cologne, 50937 Cologne, Germany}

\author{Yoni Schattner}
\affiliation{Department of Physics, Stanford University, Stanford, CA 94305, USA}
\affiliation{Stanford Institute for Materials and Energy Sciences, SLAC National Accelerator Laboratory and Stanford University, Menlo Park, CA 94025}

\author{Simon Trebst}
\affiliation{Institute for Theoretical Physics, University of Cologne, 50937 Cologne, Germany}

\author{Erez Berg}
\affiliation{Department of Condensed Matter Physics, The Weizmann Institute of Science, Rehovot, 76100, Israel}

\date{\today}

\begin{abstract}

We present numerically exact results from sign-problem free quantum Monte Carlo simulations for a spin-fermion model near an $O(3)$ symmetric antiferromagnetic (AFM) quantum critical point. We find a hierarchy of energy scales that emerges near the quantum critical point. At high energy scales, there is a broad regime characterized by Landau-damped order parameter dynamics with dynamical critical exponent $z=2$, while the fermionic excitations remain coherent. The quantum critical magnetic fluctuations are well described by Hertz-Millis theory, except for a $T^{-2}$ divergence of the static AFM susceptibility. This regime persists down to a lower energy scale, where the fermions become overdamped and concomitantly, a transition into a  $d-$wave superconducting state occurs. These findings resemble earlier results for a spin-fermion model with easy-plane AFM fluctuations of an $O(2)$ SDW order parameter, despite noticeable differences in the perturbative structure of the two theories. In the $O(3)$ case, perturbative corrections to the spin-fermion vertex are expected to dominate at an additional energy scale, below which the $z=2$ behavior breaks down, leading to a novel $z=1$ fixed point with emergent local nesting at the hot spots [Schlief \textit{et al.}, PRX 7, 021010 (2017)]. Motivated by this prediction, we also consider a variant of the model where the hot spots are nearly locally nested. Within the available temperature range in our study ($T\ge E_F/200$), we find substantial deviations from the $z=2$ Hertz-Millis behavior, but no evidence for the predicted $z=1$ criticality.

\end{abstract}


\maketitle

\section{Introduction}
\label{sec:introduction}

Quantum criticality in itinerant many-fermion systems is of great importance in the study of strongly correlated materials. Due to strong inherent correlations, many classes of materials, such as the cuprates~\cite{keimer2015quantum}, iron-pnictides~\cite{shibauchi2014quantum}, organic superconductors~\cite{lebed2008physics}, and heavy-fermion compounds~\cite{gegenwart2008quantum} host rich phase diagrams with magnetic order, non-Fermi liquid regimes, and strange metal transport. It is generally believed that quantum critical points (QCPs) at $T=0$ are playing a role in these systems~\cite{Abanov2003,Taillefer2010,Sachdev2011,Sachdev2019}. Perhaps the most prominent feature in those materials is an extended phase of high temperature superconductivity, which is generally observed in the vicinity of antiferromagnetic order. It has been proposed~\cite{Scalapino1986, Miyake1986, Monthoux1991, Millis1992, Abanov2001, Wang2013, Metlitski2010, metlitski2010instabilities} and numerically demonstrated~\cite{Schattner2016, Wang2016, SchattnerLederer2016, Dumitrescu2016, Li2017} that nearly-quantum critical fluctuations can mediate unconventional superconductivity and anomalously enhance the transition temperature $T_c$. 
However, an in-depth understanding of this universal mechanism still remains highly desirable.

Despite of numerous analytical attempts over the past decades \cite{Abanov2000, Abanov2003,Abanov2004, Metlitski2010, Hertz1976,Millis1993,Sachdev2011} few definite statements about the critical properties could be made due to the 
inherently non-perturbative nature of the interactions between gapless bosonic (magnetic) and fermionic (electronic) 
degrees of freedom. An audacious line of research \cite{Schlief2017, Lunts2017, Lee2018} carried out by 
Sung-Sik Lee and collaborators, however, has recently made notable progress in identifying a route to a controlled expansion
in an emergent control parameter to compute the properties of a strongly coupled fixed point at the $O(3)$ antiferromagnetic QCP in an {\it exact} manner. One concrete prediction of this non-perturbative analysis is a low temperature regime exhibiting a dynamical scaling exponent $z=1$ \cite{Schlief2017, Lunts2017, Lee2018}. At this fixed point, the tendency towards superconductivity is strongly suppressed. 
This general prediction calls for a complementary numerical study inspecting the hierarchy of energy scales in the vicinity of an antiferromagnetic metallic QCP.

\begin{figure}[b]
	\includegraphics[width=0.35\textwidth]{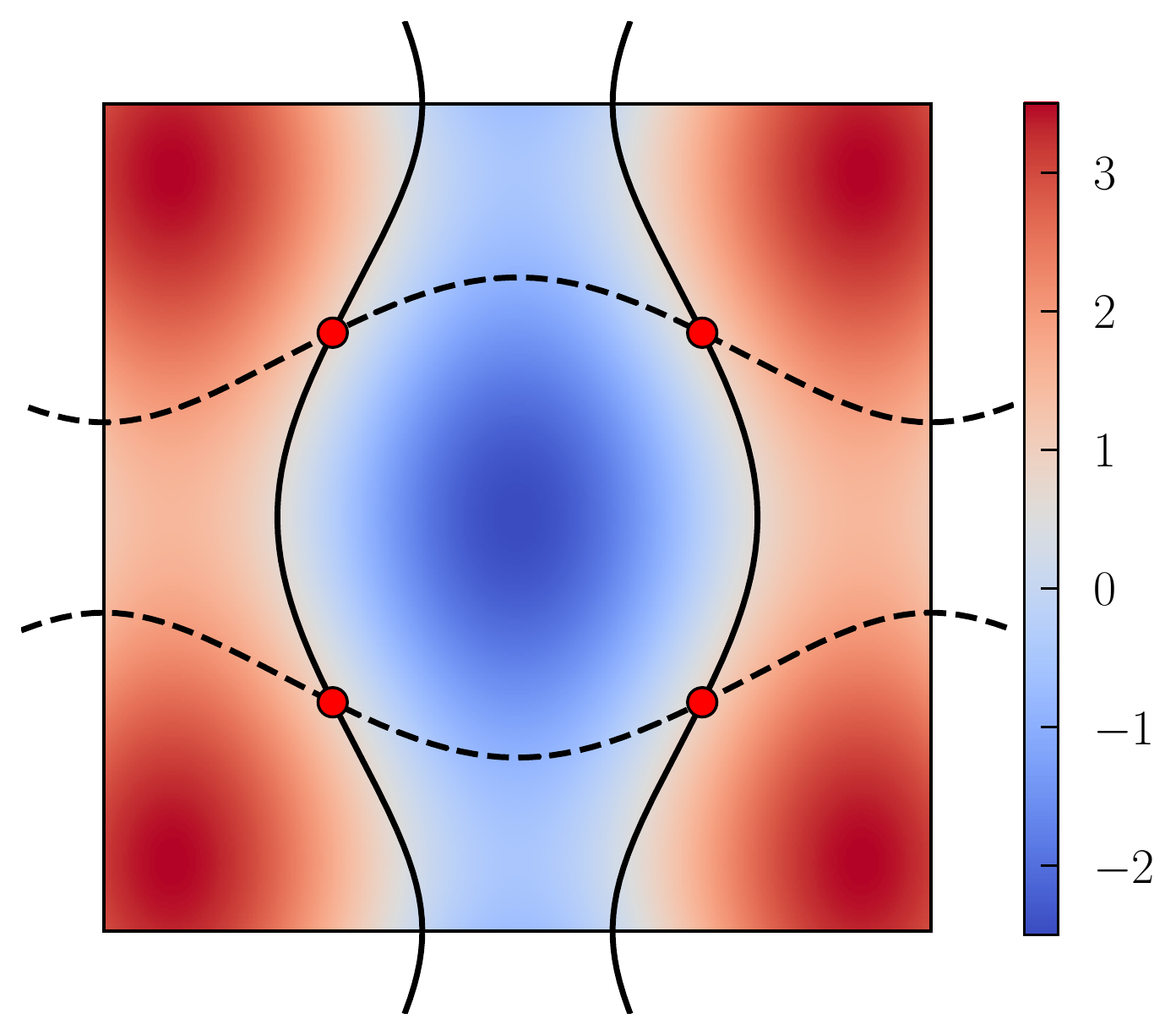}
	\caption{\textbf{Fermi surface with hot spots} across the Brillouin zone. The black lines correspond to the Fermi surfaces of the two fermion flavors $\psi_x$ (solid), $\psi_y$ (dashed). One band has been shifted by $Q=(\pi, \pi)$ such that hot spot pairs (red) occur at crossing points. The energy of the $\psi_x$ band is shown as background (color shading).}
	\label{fig:fermisurface}
\end{figure}

It is the purpose of this manuscript, to provide precisely this type of comprehensive numerical insight for an elementary microscopic model -- the spin-fermion model \cite{Hertz1976,Lee2018,Lohneysen2007} capturing the interplay of itinerant electrons and quantum critical fluctuations of an antiferromagnetic $O(3)$ spin density wave (SDW) order parameter. 
We employ large-scale quantum Monte Carlo simulations, using a controlled and unbiased Monte Carlo flavor, determinant quantum Monte Carlo (DQMC) \cite{Blankenbecler1981, Loh2005, Scalapino1993, Santos2003, Assaad2008}, which does not suffer from the infamous sign problem \cite{Berg2012}. On a technical level, it is the presence of an antiunitary symmetry \cite{Wu2005} (similar to time reversal) that renders the effective ``Boltzmann weights'' in the partition function strictly positive semidefinite. Building upon this insight, we calculate numerically exact results for the $O(3)$ antiferromagnetic QCP in a two-dimensional metal. 

Mapping out the phase diagram (shown in Fig.~\ref{fig:phasediagram} below), we find an extended region of $d$-wave superconductivity while other electronic orders, like charge density wave (CDW), do not proliferate. Above $T_c$, we find a broad temperature regime characterized by quantum critical scaling with dynamical exponent $z=2$ and overdamped dynamics of the SDW order parameter, while the fermionic excitations remain underdamped. As in the case of the recently studied easy-plane XY order \cite{Schattner2016,Gerlach2017}, the antiferromagnetic susceptibility fits a Hertz-Millis form \cite{Hertz1976,Millis1993}, except for its temperature dependence, which is consistent with $T^{-2}$. This is remarkable as it is well understood that Hertz-Millis theory is a formally uncontrolled approach \cite{Abanov2004,Metlitski2010}.
At temperatures directly above the QCP, the imaginary part of the fermionic self-energy is approximately constant at as a function of frequencies and temperatures, indicating a breakdown of Fermi liquid theory.

Finally, we search for the regime described by \citet{Schlief2017}, which is characterized by a $z=1$ dynamical scaling and strongly spatially anisotropic SDW order parameter correlations. To that end, we conduct an ultra-low temperature (down to $T \sim E_F/200$) DQMC study for a Fermi surface tuned close to local nesting at the hot spots -- a scenario that is expected to bring us closer to the  $z=1$ fixed point \cite{Schlief2017,Lunts2017}. Indeed, we observe strong deviations in the Matsubara frequency and momentum dependence of the SDW propagator from the Hertz-Millis form. 
However, we neither observe a dynamical critical exponent $z=1$ nor a momentum-anisotropy in the propagation of magnetic modes for our model parameters. This leaves us with the conclusion that, within the resolution of our DQMC simulations for systems of linear size $L \leq 14$ and temperatures $T \geq E_F/200$, we cannot numerically observe the new fixed point discovered by \citet{Schlief2017} in our $O(3)$ spin-fermion model.

Our current study expands a recent line of (predominantly) numerical efforts that have been made to extract the low-energy properties of metals at the verge of quantum phase transitions \cite{Berg2019}. Among others, QCPs related to the breaking of antiferromagnetic $Z_2$ \cite{Liu2018,Liu2018tri} and $XY$ \cite{Schattner2016, Gerlach2017}, ferromagnetic $Z_2$ \cite{Xu2017,Chubukov2009,Chubukov2004}, charge density wave \cite{Li2017}, and nematic \cite{SchattnerLederer2016, Dumitrescu2016, Li2017, Metlitski2010nematic} order have been investigated. Other recent studies focused on deconfined QCPs where fermionic matter fields are coupled to $Z_2$ \cite{Gazit2017,Chen2019,Assaad2016,Sachdev2016} and $U(1)$ \cite{Xu2019} lattice gauge theories.

Our discussion in the remainder of the manuscript is organized as follows.  In Section~\ref{sec:model} we introduce a spin-fermion model for a metal hosting an antiferromagnetic QCP. In Section~\ref{sec:theory} we provide an overview of the theoretical background from the perspective of  a perturbative expansion in the spin-fermion coupling. We discuss a hierarchy of relevant energy scales separating different physical scaling regimes and highlight, in particular, the role of vertex corrections in this context.  We then turn to an extensive DQMC analysis in Sec.~\ref{sec:dqmc} in which we map out the phase diagram of the metal for different coupling strengths before examining magnetic and metallic correlations near the critical point in detail. In Section~\ref{sec:nesting} we investigate the effect of local Fermi surface nesting on scaling properties by means of ultra-low temperature DQMC simulations. Finally, we close by discussing our findings and commenting on potential future research directions in Section~\ref{sec:discussion}.

\section{Model}
\label{sec:model}

\begin{figure}[t]
	\includegraphics[width=0.48\textwidth]{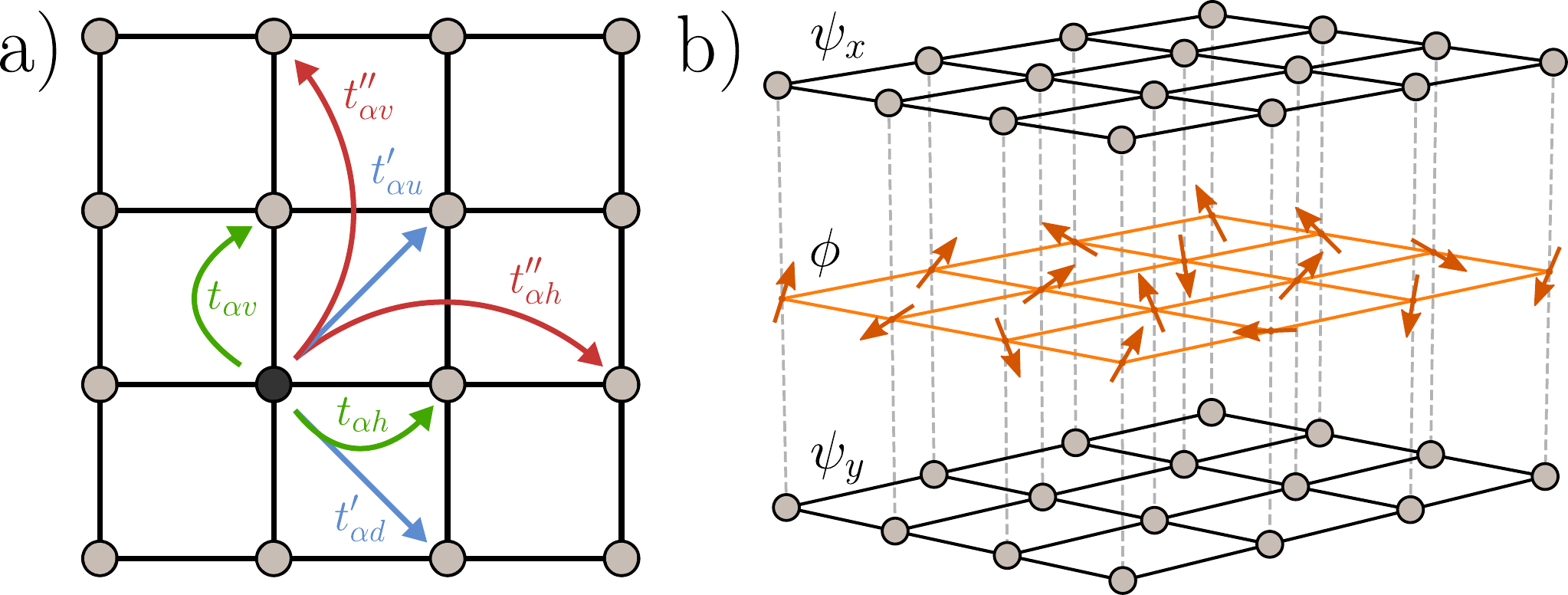}
	\caption{\textbf{Electron hopping and local interactions on the square lattice}. Illustration of nearest (green), next-nearest (blue), and next-next-nearest neighbor (red) hopping processes (a) and local Yukawa interaction (dashed lines) of the two fermion flavors $\psi_x, \psi_y$ and the magnetic order parameter $\phi$ (b) present in model \eqref{eq:model}.}
	\label{fig:square_lattice}
\end{figure}

In the vicinity of a metallic SDW quantum critical point, the relevant low-energy degrees of freedom \cite{Chubukov2002, Sachdev2011, Abanov2004} are the antiferromagnetic collective modes and electrons at the hot spots, a set of points on the Fermi surface where spin-fermion scattering is resonant, see Fig.~\ref{fig:fermisurface}.
To describe the interplay between these gapless fermionic and bosonic modes, 
we consider a low-energy effective action $S = S_\psi + S_\lambda + S_\phi$ on the two-dimensional square lattice, in which two flavors of \mbox{spin-$\frac{1}{2}$} fermions are coupled to an isotropic three-component SDW order parameter $\vec{\phi}$  which tends to order \cite{Berg2012} at a commensurate antiferromagnetic ordering wave vector $Q=~(\pi,\pi)$,
\begin{eqnarray}
S_\psi &=&  \int d\tau\sum_{ \mathbf{r}, \mathbf{r}'} \sum_{s, \alpha} \psi_{\alpha \mathbf{r}s}^\dagger \left[ \left(\partial_\tau - \mu\right)\delta_{\mathbf{r}\mathbf{r'}} - t_{\alpha \mathbf{r}\mathbf{r'}} \right]  \psi_{\alpha \mathbf{r'}s} \nonumber \,, \\
S_\lambda &=& \lambda \int d\tau\sum_{ \mathbf{r}} \sum_{s,s'} e^{i \mathbf{Q}\cdot \mathbf{r}} \vec{\phi}_\mathbf{r} \cdot \left( \psi_{x\mathbf{r}s}^\dagger \vec{\sigma}_{ss'}  \psi_{y\mathbf{r}s'} + \textrm{h.c.} \right) \label{eq:model} \,, \\
S_\phi &=& \int d\tau\sum_{ \mathbf{r}} \left[\frac{1}{2c^2} \left(\partial_\tau \vec{\phi}\right)^2 + \frac{1}{2} \left(\nabla \vec{\phi} \right)^2 + \frac{r}{2}\vec \phi^2 + \frac{u}{4} (\vec \phi^2)^2\right] \nonumber \,.
\end{eqnarray}
Here $\mathbf{r}, \mathbf{r'}$ are coordinates on the square lattice, $s \in \{\uparrow, \downarrow\}$ denotes spin, $\alpha \in \{x,y\}$ is a fermion flavor index,  $\vec{\sigma}$ is a vector of the Pauli matrices, and $\nabla$ is the lattice gradient. Imaginary time is denoted as $\tau$ and the corresponding integrals implicitly go from zero to inverse temperature $\beta = 1/T$.

The first part, $S_\psi$, describes the kinetic energy of the fermions $\psi_{\alpha s}$. SDW ordering is modelled by a canonical $\phi^4$-theory, $S_\phi$, which couples to the fermion spin density via a Yukawa coupling in $S_\lambda$. We set the bare velocity of the bosonic collective mode to $c=3$, the quartic coupling to $u=1$, and the fermion-boson Yukawa coupling to $\lambda=1$ (unless stated otherwise). The parameter $r$,  related to the mass of the order parameter field,  allows us to tune through a SDW transition. 

Specifically, unless otherwise noted, we consider nearest neighbor hopping amplitudes $t_{xh}=t_{yv}=1$, $t_{yh}=t_{xv}=0.5$ where subscripts $h, v$ indicate horizontal and vertical lattice directions, respectively. The chemical potential has opposite sign for the two fermion flavors, $\mu_x = - \mu_y = -0.5$. The resulting Fermi surface is displayed in Fig~\ref{fig:fermisurface}. Note that for this choice of parameters model \eqref{eq:model} has a two-fold rotational symmetry, consisting of a combination of a $\pi/2$ real-space rotation, shifting the momentum by $\mathbf{Q}$, a particle-hole transformation, and interchanging fermion flavors, i.e. \mbox{$\psi_{\alpha \mathbf{r} s} \rightarrow s e^{i \mathbf Q \cdot \mathbf r} \psi^\dagger_{-\alpha R_{\pi/2}(\mathbf{r}) -s}$}. This symmetry implies that hot spots occur on the momentum diagonals, $k_{x} = k_{y}$.

\section{Theoretical background} 
\label{sec:theory}

We start our discussion of the spin-fermion model \eqref{eq:model} by providing a brief overview of its relevant energy scales, summarizing some essential aspects of its theoretical background along the way.
To identify a number of crossover regimes and to understand the basic impact of the spin-fermion interactions, it is  instructive to perform an analysis for small Yukawa couplings $\lambda$ -- despite the fact that, strictly speaking, the Yukawa coupling is a relevant parameter in the sense that conventional perturbation theory becomes unreliable at sufficiently small energy scales.

\begin{figure}[t]
	\centering
	\includegraphics[width=0.45\textwidth]{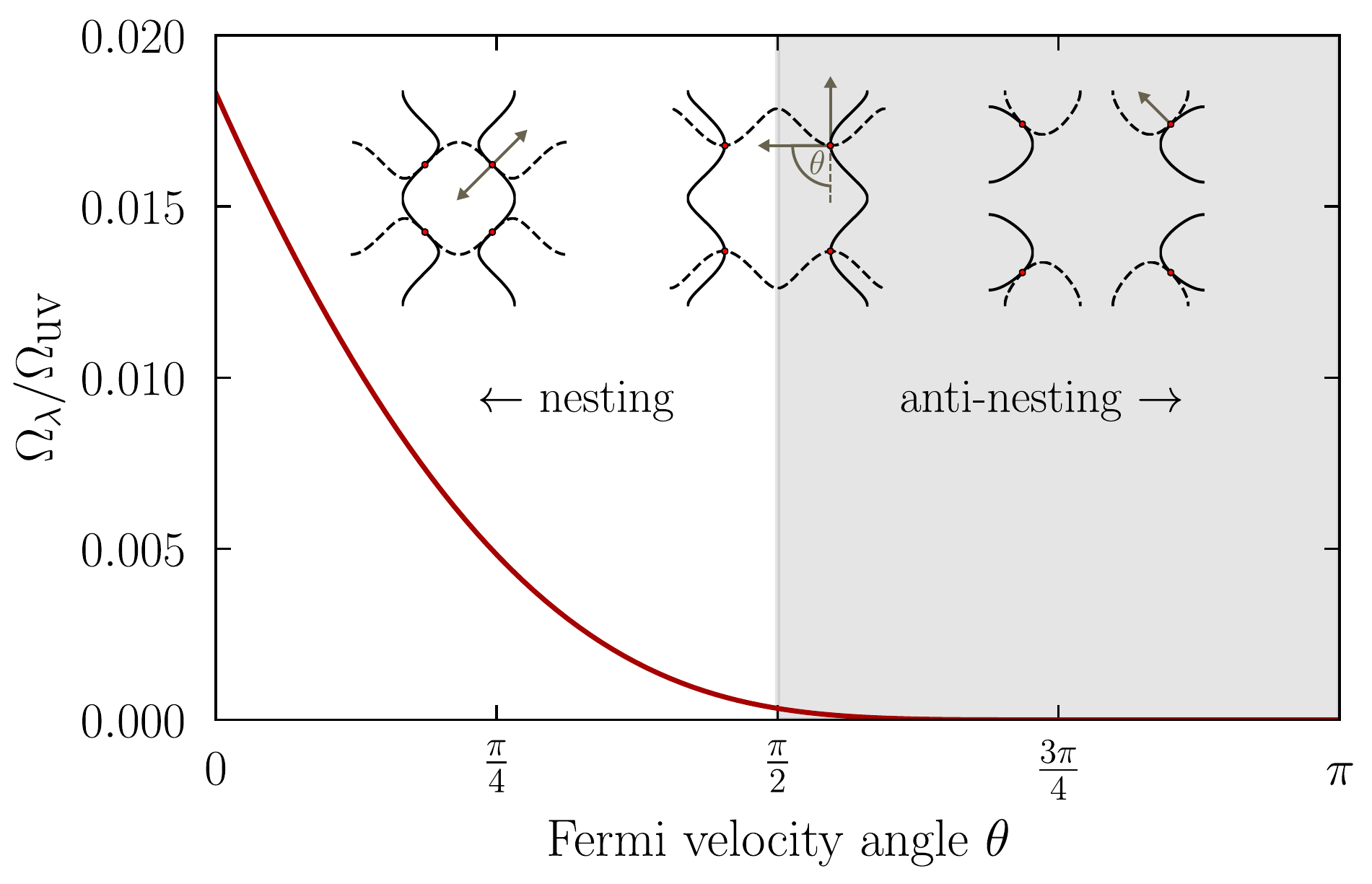}
	\caption{\textbf{Effect of nesting on the vertex energy scale} $\Omega_\lambda$ given in Eq. \eqref{eq:vertex_scale}. Here, $\theta$ is the relative angle between the Fermi velocities (as indicated in the insets). Perfect local nesting corresponds to $\theta = 0$ and anti-nesting to $\theta = \pi$.}
	\label{fig:vertexscale}
\end{figure}

Focusing on the antiferromagnetic order parameter first, we can extract an energy scale $\Omega_b$, at which the nature of the dynamics of collective spin excitations changes due to interactions with surrounding fermions. To leading order in $\lambda$,  the propagator of the bosonic field $\phi$  takes the renormalized form
\begin{equation}
\chi^{-1}(\mathbf{q},i\omega_{n})=\tilde{r}+\mathbf{q}^{2}+\frac{\omega_n^2}{c^2}+\gamma|\omega_{n}|+\ldots, \label{eq:renormprop}
\end{equation}
in which $\tilde{r}$ is the renormalized tuning parameter and
\begin{align}
	\gamma = \dfrac{N \lambda^2}{\pi v_F^2 \sin\theta} \label{eq:gamma} 
\end{align}
is a damping constant. Here, $N=4$ is the number of hot spot pairs, $\theta \in [0,\pi]$ is the angle between the Fermi velocities (of magnitude $v_F$) at the hot spots (Fig.~\ref{fig:vertexscale}). 
By estimating when the  dynamically generated contributions to the action dominate over bare ones one then finds \cite{Metlitski2010,Abanov2000, Abanov2003, Wang2016}
for the energy scale $\Omega_b$
\begin{equation}
\Omega_{b}=\frac{N\lambda^{2}c^{2}}{\pi v_F^2 \sin \theta}\,.
\end{equation}
At frequencies small compared to $\Omega_b$ the dynamics of collective magnetic modes is overdamped.
 Because of the emergence of the Landau-damping term $\gamma |\omega_n|$ in Eq.~\eqref{eq:renormprop} the dynamical critical exponent $z$ is found \cite{Metlitski2010,Abanov2000, Abanov2003, Wang2016, Sachdev2011} to increase from 1 (at the Wilson-Fisher fixed point) to 2.

Next, we consider the lowest-order effect of the renormalized bosons on the fermion dynamics at the hot spots. To leading order in  $\lambda$ one finds \cite{Metlitski2010,Abanov2000} that, at zero temperature, the fermions acquire a self-energy 
\begin{align}
\Sigma(i\omega_n) \sim \frac{in_b\lambda \sqrt{\sin{\theta}}}{\sqrt{N}}\sqrt{|\omega_n|} \,,
\end{align}
where $n_{b} = 3$ is the number of order parameter components. This implies that the fermions  become strongly damped by the coupling to SDW fluctuations, with a damping rate that scales as $\sqrt{\omega}$, indicating a distinct deviation from ordinary Fermi liquid character ($\omega^{2}\log(1/\omega)$ in two spatial dimensions). Taking the same approach as above, we estimate the energy scale $\Omega_f$ at which this breakdown occurs as
\begin{align}
	\Omega_f &\sim \frac{n_b^2\lambda^2 \sin{\theta}}{N}\,. \label{eq:omegaf}
\end{align}
For frequencies $\omega \ll \Omega_f$ the feedback from `dressed' bosons on fermion excitations at the hot spots is strong and leads to quasi-particle decoherence.
\begin{figure}[b]
	\centering
	\includegraphics[width=0.3\textwidth]{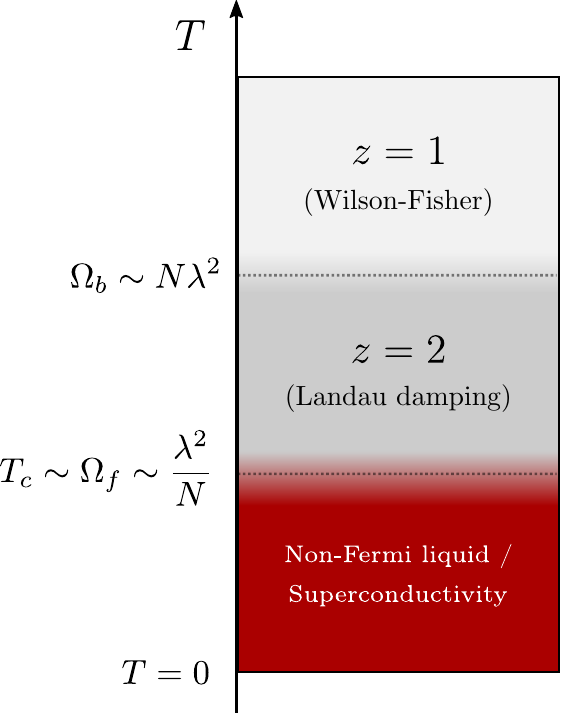}
	\caption{\textbf{Scaling regimes suggested by perturbation theory}. There is a parametrically large (in the number of hot spot pairs $N$) Landau damped regime. For the spin-fermion model \eqref{eq:model} with a Fermi surface as in Fig.~\ref{fig:fermisurface}, $N=4$.}
	\label{fig:regimes}
\end{figure}
Apart from entering a non-Fermi liquid regime, the system  may eventually become unstable against superconductivity at sufficiently low temperatures. Close to the critical point collective spin fluctuations can generate attractive pairing interactions and facilitate the formation of Cooper pairs \cite{Lee2018, Schattner2016}. At the hot spots, and for small $\lambda$, one can estimate the energy scale of the onset of superconductivity, $T_c$, by solving the Eliashberg equation for the superconducting vertex in the Landau-damping regime, where self-energy corrections are negligible. For our model \eqref{eq:model} this predicts  \cite{Wang2016} the superconducting
susceptibility to diverge at a scale
\begin{equation}
T_{c}\sim\frac{n_{b}^{2}\lambda^{2} \sin\theta}{N}\,. \label{eq:scscale}
\end{equation}
Note that the energy scales $T_c$ and $\Omega_f$ are both of the order $O(\frac{\lambda^2}{N})$ and thus there is a priori no parametrically large separation between a non-Fermi liquid and a superconducting regime.

We summarize this hierarchy of energy scales obtained from a small $\lambda$ analysis in Fig.~\ref{fig:regimes}. We note that the scales $\Omega_b$ and $\Omega_f$ are both of order $\lambda^2$ in the Yukawa coupling. This is in stark contrast to the case of an Ising-nematic QCP (where $\Omega_b \sim \lambda$ and $\Omega_f \sim \lambda^4$) which shows a parametrically large window of Landau-damped physics (with $z=3$) \cite{Berg2019, SchattnerLederer2016}. However, the energy scales are separated by their dependence on the number of hot spot pairs, $N = 4$, which may serve as a control parameter for an extended Fermi liquid regime with relaxational boson dynamics.
\begin{figure*}[t]
	\centering
	\includegraphics[width=0.47\textwidth]{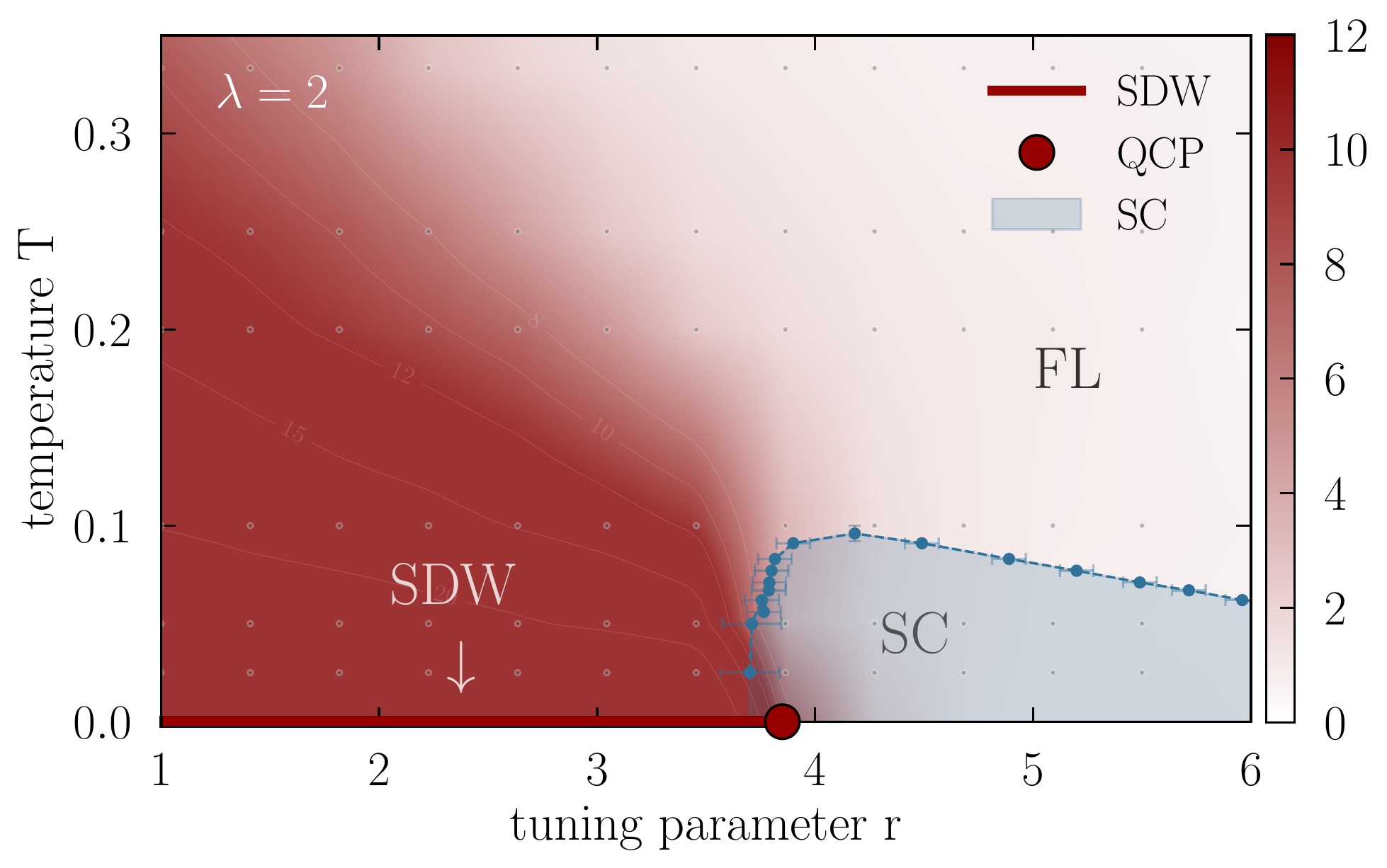}
	\hspace{2em}
	\includegraphics[width=0.47\textwidth]{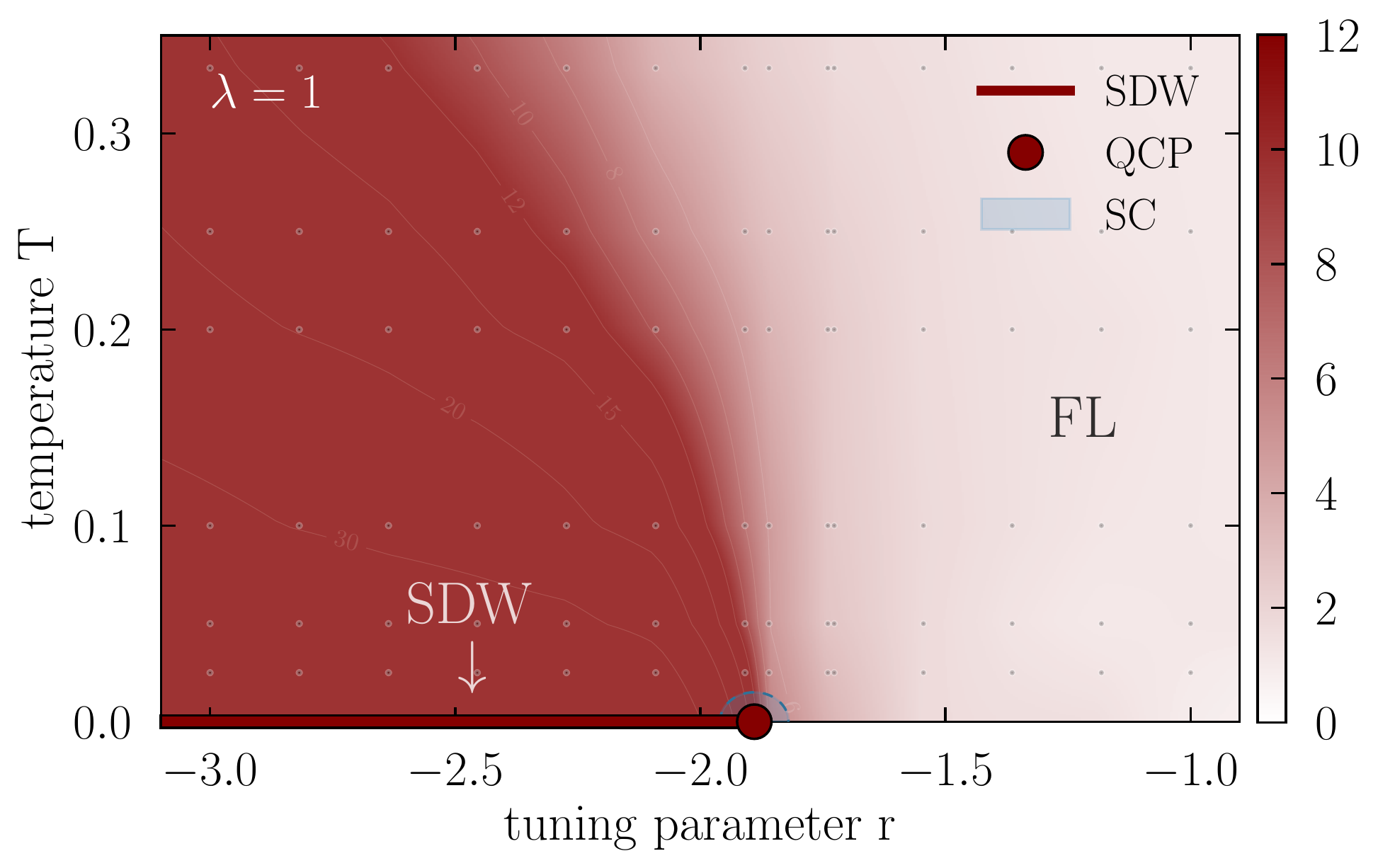}
	\caption{\textbf{Phase diagram of the spin-fermion model}, Eq.~\eqref{eq:model}, for $\lambda = 2$ (left) and $\lambda=1$ (right). The correlation length $\xi_{\textrm{AFM}}$ is shown for a $L=12$ system (color coding, paramater grid indicated by grey points). In accordance with the Mermin-Wagner theorem there is no magnetic phase transition at finite temperatures but only a crossover originating from the QCP at $r = r_c$. For large tuning parameter values $r\gg r_{c}$  the system is in an ordinary Fermi-liquid phase. In the opposite limit, $r \ll r_{c}$ the SDW correlation length diverges as $T\rightarrow0$. For $\lambda = 2$, an extended dome-shaped $d$-wave superconducting phase is masking the QCP (blue). The maximal $T_c$ is of the order of $E_F/20$. For $\lambda = 1$, the system stays non-superconducting down to the lowest considered temperature, $T=0.025$. Depending on the coupling strength, the quantum critical point, marking the onset of SDW order at $T=0$, lies at $r_{c} \approx 3.85$ ($\lambda=2$) or $r_{c} \approx -1.89$ ($\lambda=1$), respectively.
		\label{fig:phasediagram}}
\end{figure*}

Finally, we inspect first-order corrections $\delta \Gamma$ to the spin-fermion vertex itself. Following a similar procedure, one can extract an energy scale $\Omega_\lambda$ at which the dynamically generated higher-order interactions becomes comparable  to the bare vertex, $\delta \Gamma \sim \lambda$. One obtains \cite{Metlitski2010}
\begin{align}
\Omega_{\lambda} = \Omega_{\textrm{uv}}\, \exp{\left( -\frac{\pi N}{\pi - \theta} \right)}\,, 
\label{eq:vertex_scale}
\end{align}
where $\Omega_{\textrm{uv}}$ is an ultra-violet cutoff and $0< \theta< \pi$. Note $\Omega_\lambda$ does not depend on the coupling strength $\lambda$ -- the naive $\lambda^{2}$ is compensated by the Landau-damping constant $\gamma$. Instead, for a fixed number of hot spot pairs, the scale $\Omega_\lambda$ is only sensitive to the relative angle $\theta$ between the Fermi surfaces at the hot spots. We illustrate the dependence of $\Omega_\lambda$ on $\theta$ in Fig.~\ref{fig:vertexscale}. In particular, we see a moderate increase of $\Omega_\lambda$ in the limit of (local) nesting, $\theta \rightarrow 0$.
Let us remark that Eq.~\eqref{eq:vertex_scale} is special to the case of an $O(3)$ symmetric SDW order parameter. Generically, the leading order correction to the boson-fermion vertex scales as $\delta \Gamma \sim 2- n_b $, and thus precisely vanishes for easy-plane antiferromagnetism.

Despite the general guidance provided by such a perturbative treatment of the Yukawa coupling, this approach is formally uncontrolled and must be complemented by a more sophisticated analysis. Many intensive analytical efforts have been made to get a handle on strong interactions and gain control over calculations. These include various $1/N$-expansions \cite{Metlitski2010, Abanov2000, Abanov2003} and extensions of the problem to fractional dimensions \cite{Sur2015, Lunts2017}. One of the more striking developments has been a recent, self-consistent and non-perturbative study which has been put forward by \citet{Schlief2017}. It utilizes an emergent control parameter -- the degree of local nesting at the hot spots -- to compute  critical exponents in an exact manner. Notably, the identified strong coupling fixed point is characterized by the following form of the SDW susceptibility: 
\begin{equation}
\chi^{-1}\sim \gamma' |\omega_n| + |q_x + q_y| + |q_x - q_y| \,,
\end{equation}
where $\gamma'$ is a parameter.
Note that the SDW correlations are highly anisotropic in space, and have a dynamical exponent $z=1$.  
In this regime, the fermions remain underdamped. 

Clearly, a confirmation of those predictions with an unbiased method would be highly desirable and the motivation for the numerical studies to be reported in the following.
A central open question is whether the revealed $z=1$ fixed point is a generic property of a SDW QCP, in the sense of an extensive basin of attraction, or rather limited to a narrow parameter range.

\section{Determinant quantum Monte Carlo}
\label{sec:dqmc}
We investigate the physics of model~\eqref{eq:model} by means of extensive finite-temperature DQMC \cite{Blankenbecler1981, Loh2005, Scalapino1993, Santos2003, Assaad2008} simulations. The key feature of these quantum Monte Carlo methods is that they are numerically exact - given sufficient computation time both the statistical Monte Carlo error and the systematic Trotter error can be made arbitrarily small - and do not rely on the smallness of any expansion parameter. Provided the absence of the famous sign-problem, they allow us to explore the relevant region of the exponentially large configuration space in polynomial time. For our two-flavor spin-fermion model, it has been proven that probability weights are strictly positive because of the presence of an antiunitary symmetry similar to time reversal \cite{Berg2012}.

\subsection{Phase diagram}
\label{sec:phasediagram}
Before turning to quantum critical properties, we show the phase diagram of model~\eqref{eq:model} for coupling, $\lambda = 2$, in Fig.~\ref{fig:phasediagram}.
In two dimensions, the $O(3)$ symmetry present in $S_\phi$ cannot be spontaneously broken at finite temperatures according to Mermin and Wagner's theorem \cite{Mermin1966}. Magnetic correlations, indicated by the SDW susceptibility
\begin{align}
\chi_0(\mathbf{q}) &= \int_\tau \sum_{\mathbf r} e^{- i\mathbf{q} \cdot \mathbf{r}} \left\langle \vec{\phi}_{\mathbf{r}}(\tau) \cdot \vec{\phi}_0(0) \right\rangle,
\end{align}
in which we measure momenta relative to the antiferromagnetic ordering wave vector $\mathbf{Q} = (\pi, \pi)$, are therefore bound to be finite-ranged at any nonzero $T$ and there is no classical phase transition originating from the QCP at $T=0$. Anticipating that, at large length scales, the susceptibility is of Ornstein-Zernike form \cite{OrnsteinL.S.Zernike1918},
we can define a correlation length of antiferromagnetic fluctuations \cite{OrnsteinL.S.Zernike1918, Chaikin1996}
\begin{align}
    \xi_{\textrm{AFM}} = \dfrac{L}{2\pi}\sqrt{\dfrac{\chi(q=0)}{\chi(q=2\pi/L)} - 1} \,,
\end{align}
with $q=2\pi/L$, the smallest non-vanishing momentum of a finite system. In Fig.~\ref{fig:phasediagram} we display $\xi_{\textrm{AFM}}$ across the $(r,T)$-plane using a linear interpolation between a discrete set of $r$ values and temperatures (indicated by dots in the figure). As can be clearly seen, the data manifestly reveals a finite-temperature magnetic crossover culminating into a quantum critical point marking the onset of SDW order at $T=0$.

\begin{figure}[b]
	\centering
	\includegraphics[width=0.47\textwidth]{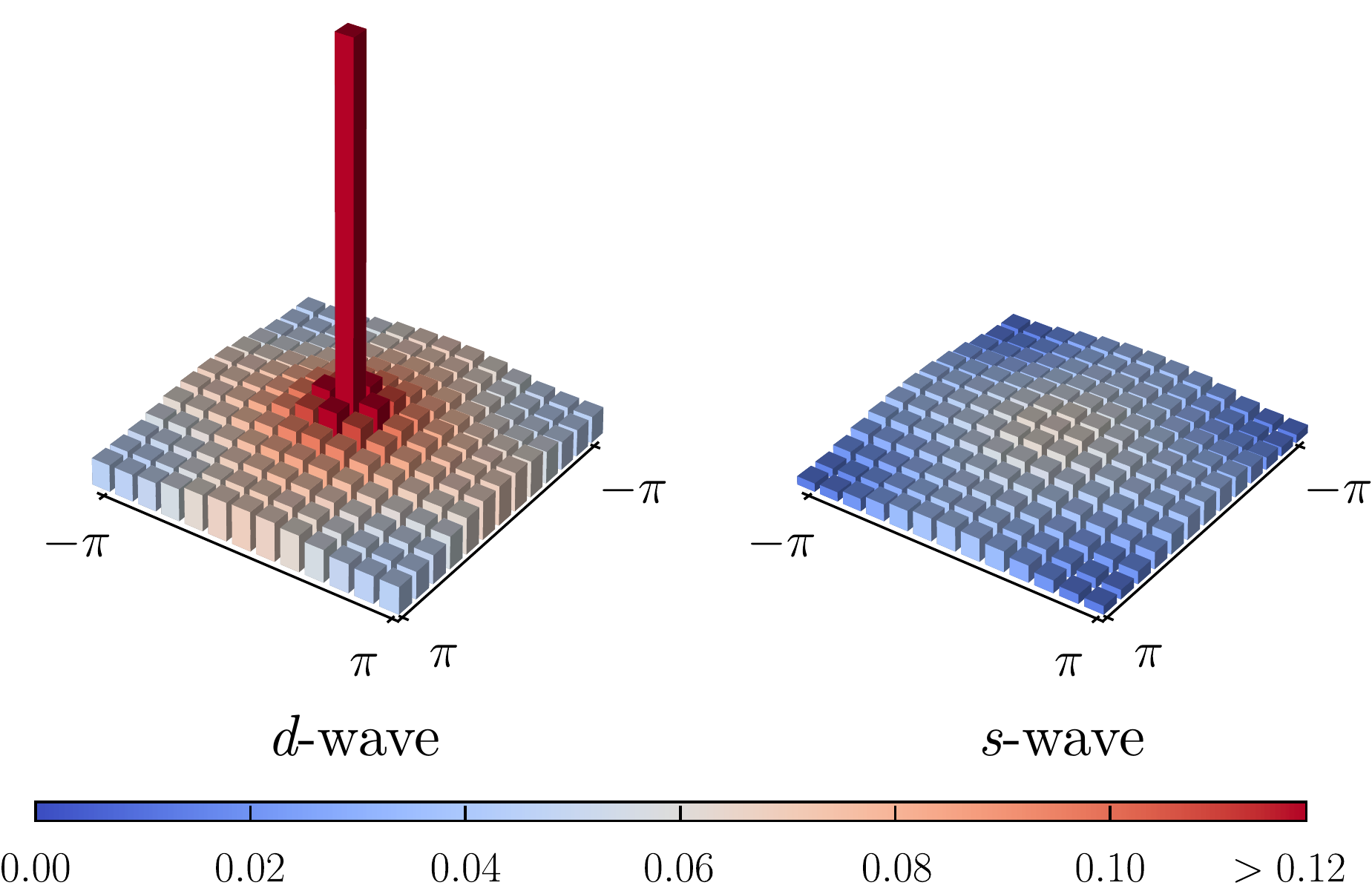}
	\caption{\textbf{Nature of the superconducting state}. Momentum resolved equal-time pairing correlations across the first Brillouin zone with $d$-wave (left) and $s$-wave symmetry (right) close to the quantum critical point for $\lambda=2$, $\beta~=~20$, and $L=12$. While the $s$-wave correlations are featureless there is a distinct peak in the $d$-wave channel.}
	\label{fig:etpc_momentum_3d}
\end{figure}

While the Mermin and Wagner theorem also holds for the continuous $U(1)$ symmetry associated with superconductivity there is the possibility of a Berezinskii-Kosterlitz-Thouless (BKT) transition for the two dimensional superconducting order parameter. We determine the superconducting transition temperature $T_c(r)$ as the temperature where the superfluid density of the system surpasses the universal BKT value $\Delta\rho_{s} = 2T/\pi$ \cite{Scalapino1993, Schattner2016}. The resulting phase transition is shown in Fig.~\ref{fig:phasediagram}. For $\lambda = 2$, we observe an extended dome-shaped superconducting phase that shows its peak critical temperature $T^{\textrm{max}}_c \approx 0.09$ close to the QCP. We find that the value for $T^{\textrm{max}}_c$ is in good agreement with an independent self-consistent Eliashberg calculation which gives $T^{\textrm{El}}_c \approx 0.08$. 
The nature of the superconducting state reveals itself in the uniform pairing susceptibility $P_{\eta} = \int d\tau \sum_i \langle \Delta_{\eta}^\dagger(r_i, \tau) \Delta_{\eta}(0,0) \rangle$, with $\Delta_{\eta}(r_i) = \psi_{xi\uparrow}^\dagger \psi_{xi\downarrow}^\dagger - \eta \psi_{yi\uparrow}^\dagger \psi_{yi\downarrow}^\dagger$ and $\eta = \pm 1$. Note that under a $\pi/2$ rotation $\Delta_\eta \rightarrow \eta \Delta^\dagger_\eta$. Hence, $\Delta_{-}$ has $d$-wave character while $\Delta_{+}$ is of $s$-wave nature. In Fig.~\ref{fig:etpc_momentum_3d} we show the momentum resolved pairing susceptibilities $P_\eta$ across the first Brillouin zone. The strong peak in the sign-changing symmetry channel is evidence for a $d$-wave superconducting state.
~

In principle, quantum critical fluctuations can also promote other electronic orders like CDW formation \cite{metlitski2010instabilities,Efetov2013, Wang2014}. For particle-hole symmetric electron dispersions, $d$-wave superconductivity and $d$-wave CDW are degenerate due to an emergent $SU(2)$ symmetry \cite{Wang2018}. Inspecting the relevant susceptibilities (see App.~\ref{sec:cdc}), we find that CDW correlations are mildly enhanced at low temperatures, most strongly in the vicinity of the QCP, but are only weakly dependent on system-size. Interestingly, this amplification seems to be irrespective of the onset of superconductivity at $T=T_c$, indicating the absence of a competition between CDW and SC.

\subsection{Quantum critical correlations} 
\label{sec:magnetic_corr}
To extract the critical properties of model~\eqref{eq:model} we suppress the superconducting transition by studying the system for smaller Yukawa coupling $\lambda=1$ where the critical temperature $T_c < 1/40$ is below the lowest accessible temperatures. Simulations are then performed in the vicinity of the QCP and above any possible ultralow-temperature superconducting phase.

\subsubsection{Magnetic correlations}

We investigate the system's magnetic correlations by analyzing the bosonic SDW susceptibility
\begin{align}
	\chi(\mathbf{q},i\omega_n; r, T) &= \int_\tau \sum_{\mathbf r} e^{i\omega_n \tau - i\mathbf{q} \cdot \mathbf{r}} \left\langle \vec{\phi}_{\mathbf{r}}(\tau) \cdot \vec{\phi}_0(0) \right\rangle \label{eq:suscept}
\end{align}
at momenta $\mathbf{q}$, taken relative to the ordering wave vector $\mathbf{Q}$, and Matsubara frequencies $\omega_n = 2\pi nT$.
~
\begin{figure}
	\centering
	\includegraphics[width=0.47\textwidth]{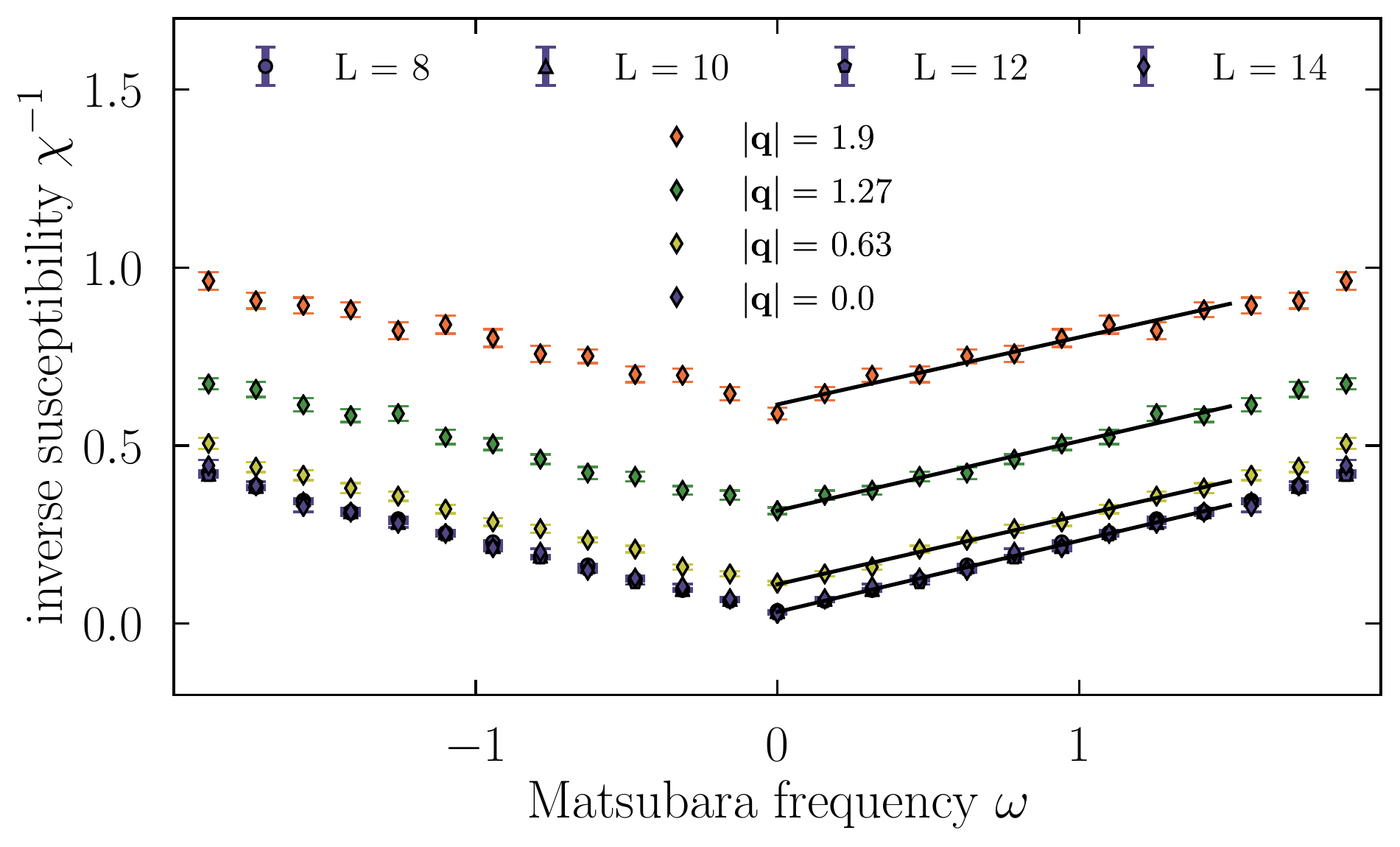}
	\caption{\textbf{Frequency dependence} of the magnetic susceptibility $\chi$ close to the quantum critical point ($r=-1.74$) at inverse temperature $\beta=40$ and $\lambda = 1.0$. The different colors indicate different momenta. The solid lines are linear low-frequency fits.}
	\label{fig:chi_omega}
\end{figure}

First, the dependence of the inverse susceptibility $\chi^{-1}(\mathbf{q}, i\omega_n)$ on Matsubara frequency in the vicinity of the QCP is illustrated in Fig.~\ref{fig:chi_omega}. The data very visibly has linear character for small Matsubara frequencies $\omega_n$, both for $\mathbf{q} = 0$ and small finite momenta $\mathbf{q} > 0$, with an apparent cusp at $\omega_n = 0$. To establish the presence of a $|\omega_n|$-term in $\chi^{-1}$ we perform a linear regression for the small frequency data. The resulting fits to the form $a_1|\omega_n| + a_0$, displayed in Fig.~\ref{fig:chi_omega} as black lines, are in good agreement with the data and confirm the linear Matsubara frequency dependence. This finding suggests overdamped dynamics of the boson field $\phi$ due to interactions with the fermions. At finite Matsubara frequencies, finite-size effects are negligibly small (within error bars), as evident in the data collapse of $\chi^{-1}(\mathbf{q} = 0, i\omega_n)$ for different system sizes, $8 \leq L \leq 14$.

\begin{figure}
	\centering
	\includegraphics[width=0.47\textwidth]{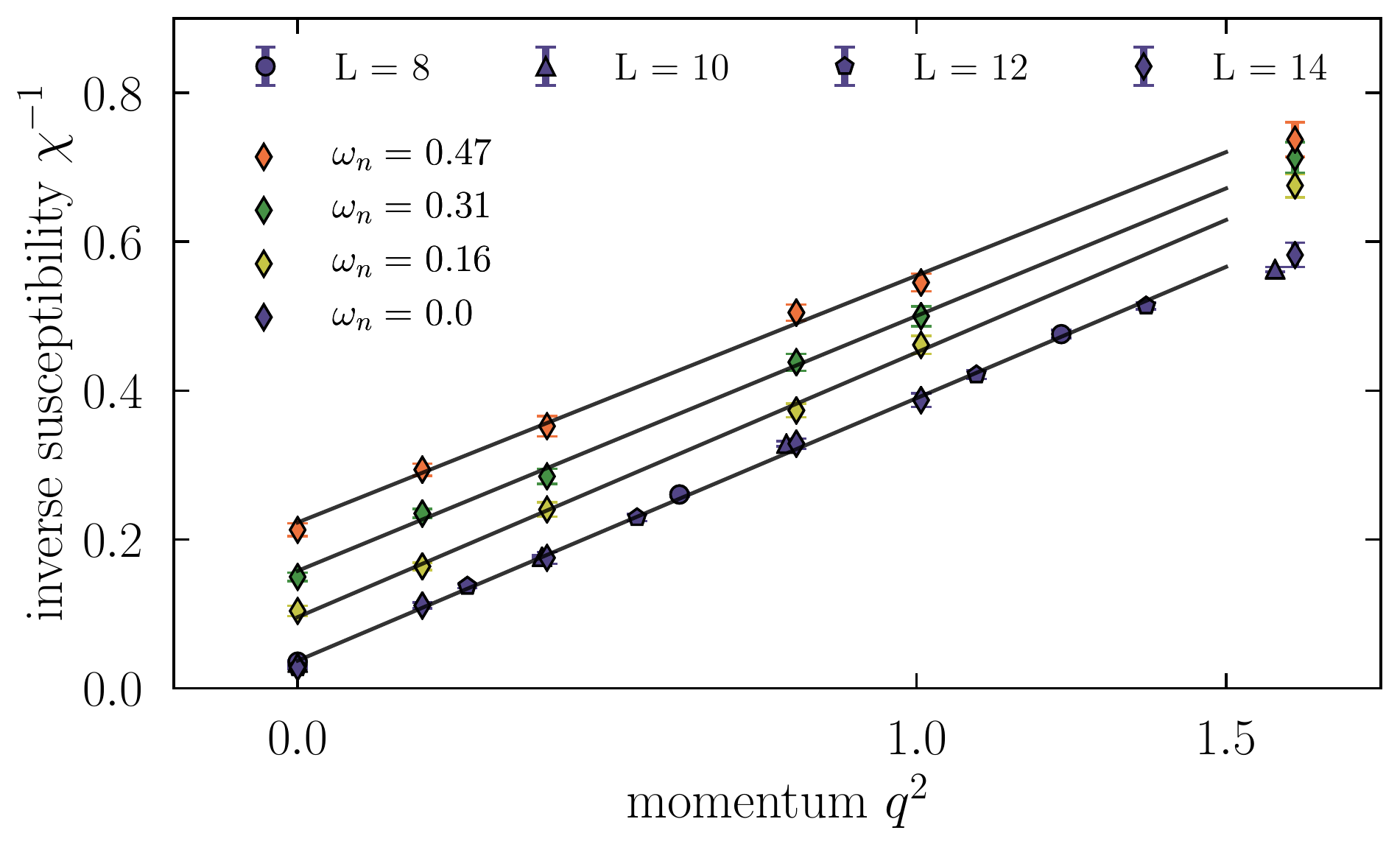}
	\caption{\textbf{Momentum dependence} of the magnetic susceptibility $\chi$ close to the quantum critical point ($r=-1.74$) at inverse temperature $\beta=40$ and $\lambda = 1.0$. The different colors indicate the lowest few Matsubara frequencies. The solid lines are linear fits.}
	\label{fig:chi_q}
\end{figure}

Turning to the momentum dependence of $\chi^{-1}(\mathbf{q}, i\omega_n)$ next, we find that the momentum dependence shown in Fig.~\ref{fig:chi_q} is consistent with a quadratic form $\mathbf{q}^2$ for small momenta $\mathbf{q}$. This holds both for $\omega_n = 0$ and small non-vanishing Matsubara frequencies $\omega_n$. Similar to above we establish the presence of a $q^2$ term in $\chi^{-1}(\mathbf{q}, i\omega_n)$ by fitting the data to the form $a_1 \mathbf{q}^2 + a_0$. The results are indicated in Fig.~\ref{fig:chi_q} as black lines, showing good agreement with the DQMC data. In combination with the observed linear Matsubara frequency dependence, this provides strong evidence for a dynamical critical exponent $z = 2$.

\begin{figure}
	\centering
	\includegraphics[width=0.47\textwidth]{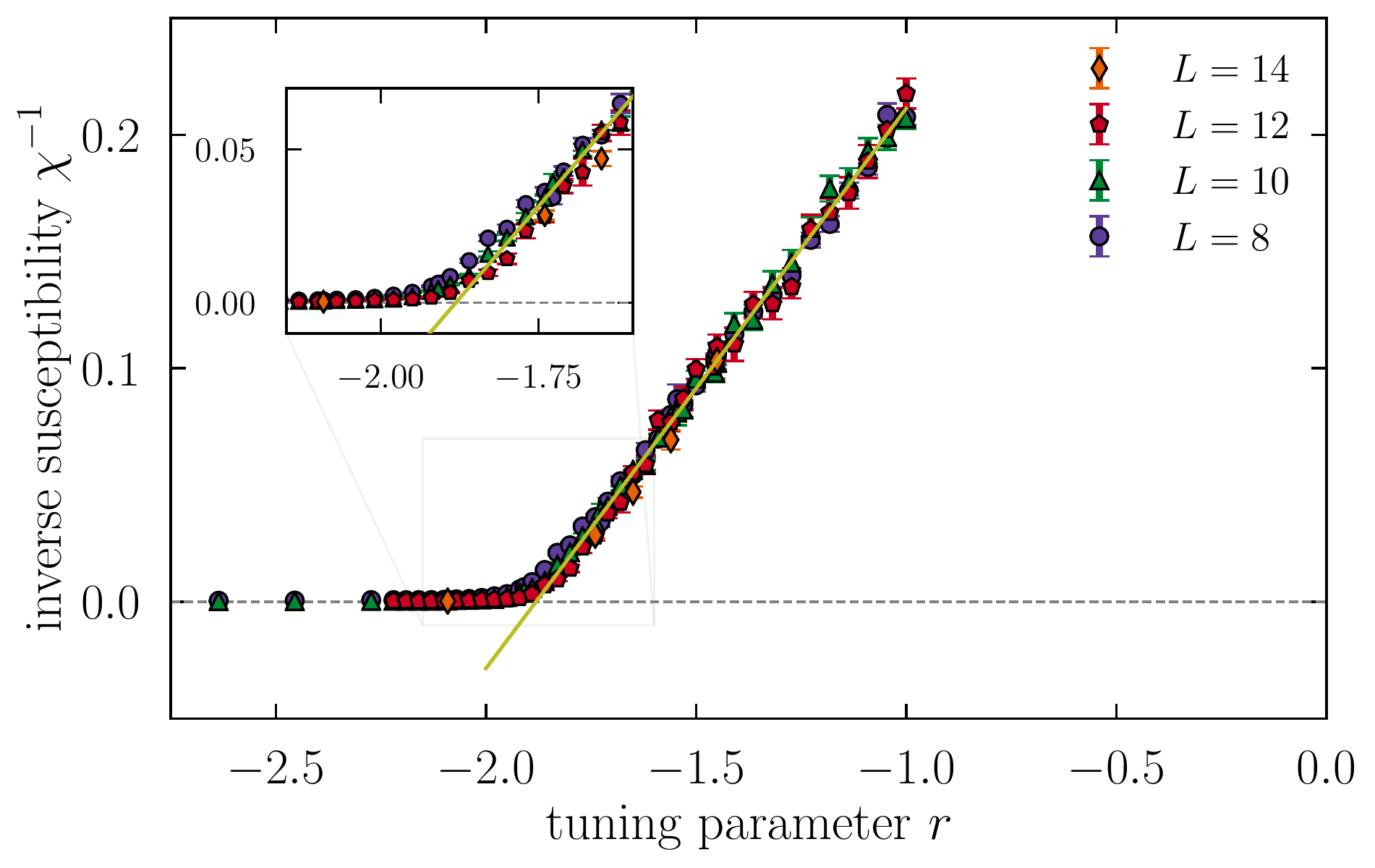}
	\caption{\textbf{Tuning parameter dependence} of the inverse magnetic susceptibility, Eq.~\eqref{eq:suscept}, close to the quantum critical point at inverse temperature $\beta=40$  and $\lambda = 1.0$. We show a linear fit of the data (light green line) with root $r_c \approx -1.89$, an estimate for the location of the QCP. \label{fig:chi_r}}
\end{figure}
~
Next, we illustrate the dependence of the inverse susceptibility $\chi^{-1}(\mathbf{q}=0, i\omega_n=0)$ on the tuning parameter~$r$ in Fig.~\ref{fig:chi_r}. For tuning parameter values~$r \geq r_0 \approx -1.89$ we find that the data for different system sizes follows a linear dependence. Due to finite-size and finite-temperature effects the onset of a finite value of $\chi^{-1}$ does not appear to be as abrupt but instead is moderately continuous \cite{Gerlach2017}.

Finally, we inspect the temperature dependence $\chi^{-1}(T)$ close to the critical point in Fig.~{\ref{fig:chi_T}. Within the limits of our numerical accuracy, we find that the DQMC results are consistent with a $T^2$ term at low temperatures. We highlight this point by fitting the data to a second degree polynomial (black line) which is able to adequately capture the temperature trend over a broad range $0.025 < T \lesssim 0.5$. At higher temperatures, the situation changes and we find a linear $T$-dependence, as shown in the inset of Fig.~\ref{fig:chi_T}.

Taken together, these findings indicate that the inverse SDW susceptibility near the metallic QCP (at low temperatures) has the form,
\begin{align}
\chi^{-1}(\mathbf{q}, i\omega_n; r, T\rightarrow 0) = c_\omega |\omega_n| + c_q \mathbf{q}^2 + c_r (r-r_0). \label{eq:HertzMillis}
\end{align}
Remarkably, this is precisely the functional dependence predicted by Hertz-Millis theory, in which fermion degrees of freedom are integrated out -- despite their gapless nature -- to obtain an effective critical theory for $\phi$ \cite{Hertz1976, Millis1993}. However, our DQMC simulations also suggest certain deviations from Hertz-Millis theory, specifically in terms of the observed quadratic temperature dependence (in lieu of a linear scaling).
Interestingly, this general comparison to Hertz-Millis theory is in full analogy to what has been observed for an $O(2)$ order parameter in Refs.~\cite{Schattner2016, Gerlach2017}.

Let us close this section by noting that it is always a possibility that the character of the onset of magnetism is first order rather than continuous. When we turn off the quartic boson coupling in model \eqref{eq:model}, i.e. $u=0$, we indeed observe a region of {\it coexistence} of magnetic and non-magnetic states. This is indicated by a jump-like tuning parameter dependence (App.~\ref{app:firstorder}) of the magnetic susceptibility $\chi^{-1}(\mathbf{q}=0, i\omega_n = 0)$ and an apparent double-peak structure in its distribution function. Keeping $u$ finite and positive should, however, generally favor a continuous transition.

\begin{figure}
	\centering
	\includegraphics[width=0.47\textwidth]{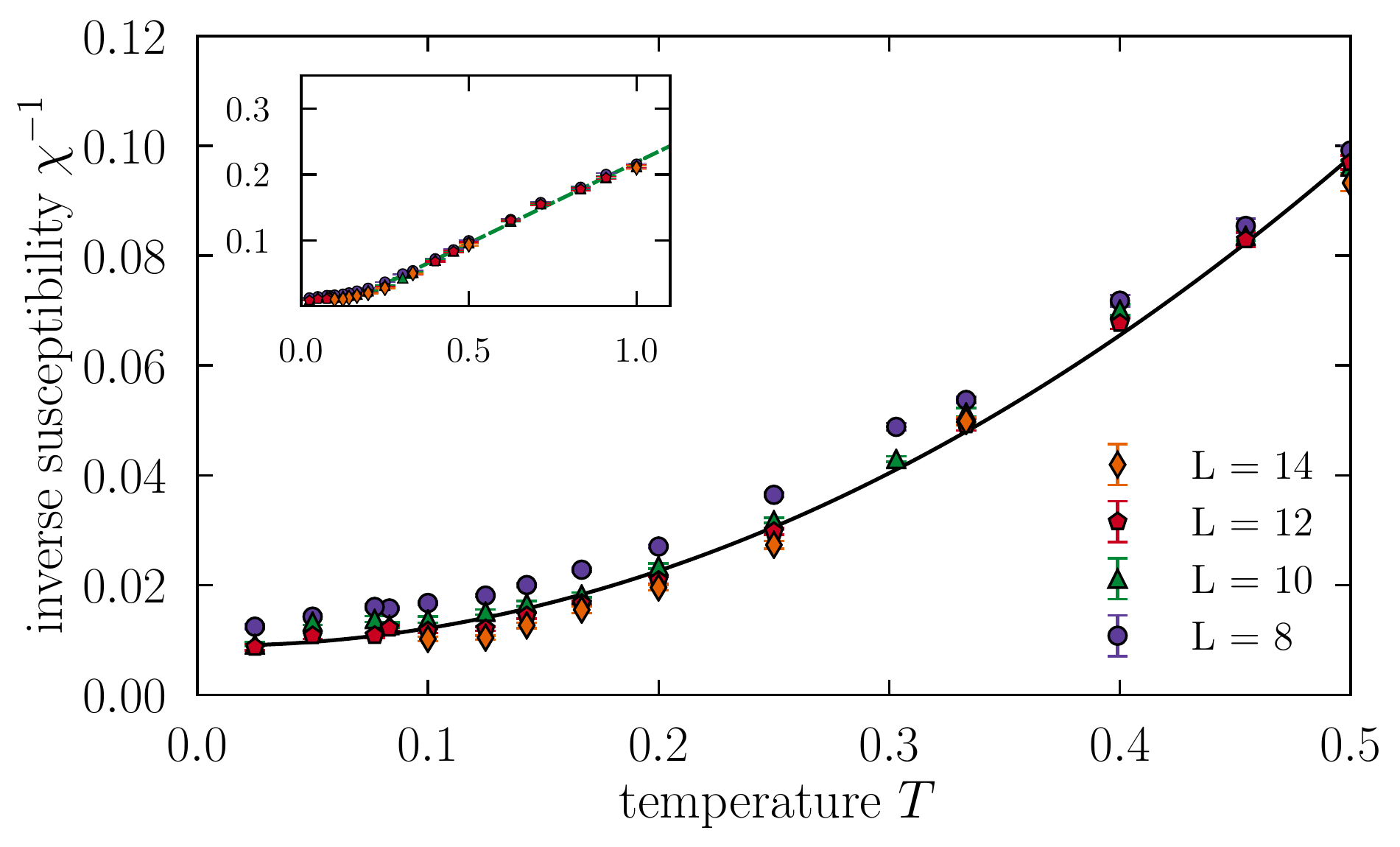}
	\caption{\textbf{Temperature dependence} of the magnetic susceptibility $\chi$ at $r \approx -1.86$, close to the quantum critical point, for $\lambda = 1.0$. The black line is a quadratic fit, $f(x) = a_2 x^2 + a_1 x + a_0$ with $a_2 \approx 0.366$, $a_1 \approx -0.005$, and $a_0 \approx 0.009$, for system sizes $L>8$.}
	\label{fig:chi_T}
\end{figure}

\subsubsection{Single fermion correlations}
\label{sec:fermioncorrelations}

\begin{figure}[t]
	\centering
	\includegraphics[width=\columnwidth]{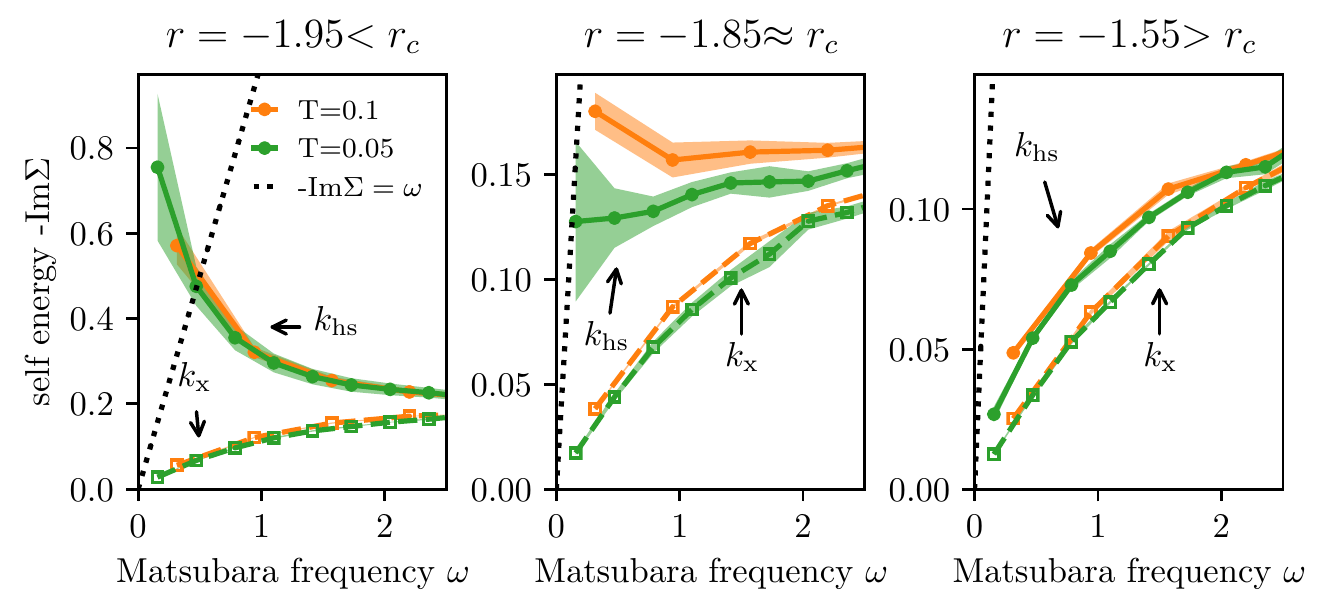}
	\caption{Imaginary part of the {\bf fermionic self-energy} as a function of Matsubara frequencies for $\lambda=1$ and $L=12$. The dotted line indicates the non-Fermi-liquid crossover scale $-\mathrm{Im}\Sigma(\omega_n) = \omega_n$. The solid lines represent the self-energy at the hotspots, $\mathbf{k}=\mathbf{k}_\mathrm{hs}$, and solid lines represent the self energy at Fermi point at $k_y=0$, namely $\Sigma(\mathbf{k}=\mathbf{k_x}=(k_f,0)$. The data points are averaged over different twisted boundary conditions, while the width of the line is the standard deviation of $\mathrm{Im} \Sigma$ between the different boundary conditions, thus representing the uncertainty due to finite-size effects, see text.
	}
	\label{fig:self_energy}
\end{figure}
Upon approaching the QCP from the disordered phase, the fermions at the hot spots lose their coherence. This is manifested in a substantial growth of the imaginary part of the (Matsubara) self-energy, defined as $\Sigma_k(\omega_n)= i\omega_n - \epsilon_k -G^{-1}_k(\omega_n)$, shown in Fig.~\ref{fig:self_energy}. At the QCP, $r\approx r_c$, the self-energy at the hotspots is very weakly dependent on both temperature and frequency. We identify the lowest temperature studied here ($T=0.05$) as the non-Fermi-liquid crossover temperature, at which the self-energy begins to dominate the bare frequency dependence of the Green's function, namely $\mathrm{Im} \Sigma_k(\omega_n=\pi T) = \pi T$.  For $r<r_c$ the imaginary part of the self-energy tends to diverge, indicating the gapping out of the hot spots. The self energy away from the hotspots remains small, and tends to vanish at low frequencies.

On a technical note, we mention that the leading cause of uncertainty in the numerical estimation of the self-energy are finite size effects. To minimize these effects we used data from several sets of twisted boundary conditions, as explained in more detail in 
Appendix \ref{sec:flux}. The time discretization error in $G_k(\omega_n)$ is reduced by using the `Filon-Trapezoidal' rule~\cite{filon_trapezoid}. 

\section{Local nesting}\label{sec:nesting}

The recent non-perturbative calculations by \citet{Schlief2017} revealed a $z=1$ fixed point with characteristics very much different from our findings above. As an attempt to promote a flow to this self-consistent critical theory, we modify our original spin-fermion model to host a Fermi surface with (almost) perfect local nesting near the hot spots. Some of the low-energy properties of this system close to criticality are then extracted in ultra-low temperature DQMC simulations.

Specifically, we introduce higher order hopping processes into Eq.~\eqref{eq:model} and choose ${t_{xh} = 1 = t_{yv}}$, ${t''_{xv} = 0.45 = -t''_{yh}}$, where $t''_{xh}$ and $t''_{yv}$ are second neighbor hopping amplitudes for $\psi_x$ in horizontal and $\psi_y$ in vertical direction, respectively (Fig.~\ref{fig:square_lattice}). All other hopping amplitudes, as well as the quartic boson coupling $u$ are set to zero, whereas $\mu_x = -\mu_y=-0.47$. The resulting Fermi surface is illustrated in Fig.~\ref{fig:fermisurface_nested}. Note that the relative angle between the Fermi velocities for this set of parameters is $\theta \approx 8.5^\circ$ whereas the Fermi surface in Fig.~\ref{fig:fermisurface} has $\theta \approx 36.9^\circ$ (the magnitude of $v_F$ does not change significantly).

\begin{figure}[t]
	\centering
	\includegraphics[width=0.35\textwidth]{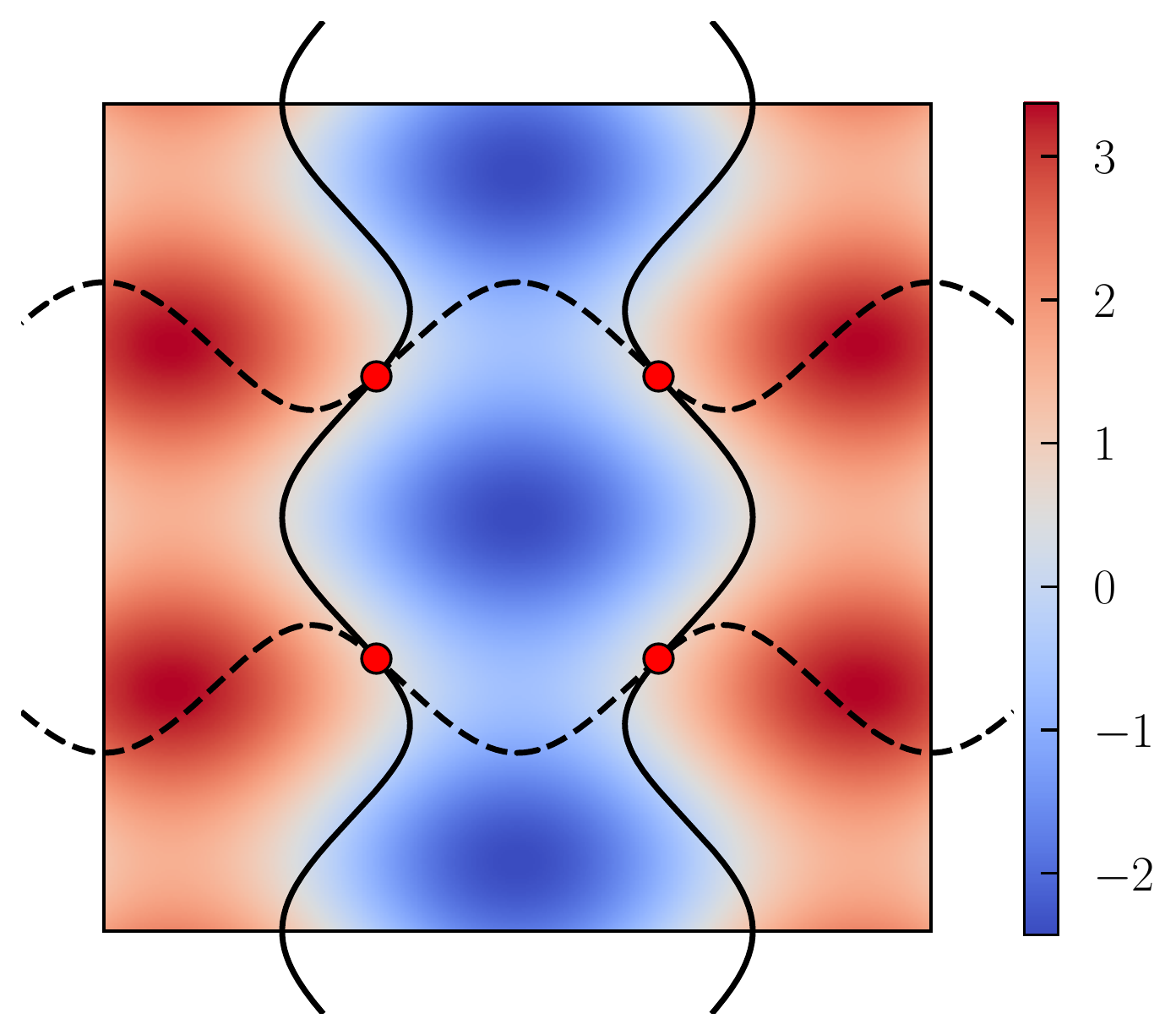}
	\caption{\textbf{Fermi surface close to local nesting near the hot spots} across the Brillouin zone. The black lines correspond to the two fermion flavors $\psi_x$ (solid), $\psi_y$ (dashed). One band has been shifted by $Q=(\pi, \pi)$ such that hot spot pairs (red) occur at crossing points.}
	\label{fig:fermisurface_nested}
\end{figure}

We start off by discussing a potential superconducting transition. In an almost locally nested system antiferromagnetic fluctuations will only slightly enhance $T_c$ \cite{Schlief2017,Wang2016,Abanov2003}. As indicated in Eq.~\eqref{eq:scscale}, the energy scale associated with superconductivity is expected to decrease as a function of the angle, $T_c \sim \sin{\theta}$. This relation has been numerically confirmed for a generic Fermi surface coupled to an $O(2)$ order parameter in Refs.~\cite{Wang2016,Gerlach2017}.  Anticipating that superconductivity arising from fermions at the hotspots, Eq.~\ref{eq:scscale}, is independent of the precise band structure \cite{Wang2016} we expect a fourfold reduction of $T_c$ for the Fermi surface shown in Fig.~\ref{fig:fermisurface_nested} compared to the one considered in Sec.~\ref{sec:model}. Indeed, we do not observe superconductivity in our DQMC simulations for temperatures as low as $T_c = 0.01$.

In Fig.~\ref{fig:nested_chi_omega}, we show the Matsubara frequency dependence of the inverse SDW susceptibility $\chi^{-1}(\mathbf{q} = 0, i\omega_n)$. In the interval $0< \omega_n < 1$ a distinct curvature can be seen, which is more pronounced at smaller Matsubara frequencies. To resolve this nonlinearity, simulations have been performed at ultra-low temperatures, $\beta = 100$. For $\omega_n > 0$, data points obtained from DQMC simulations of differently sized systems fall on top of each other, indicating the absence of significant finite size effects. To understand the origin of the curvature, we consider the non-interacting fermionic susceptibility $\Pi_0$. As illustrated in Fig.~\ref{fig:nested_chi_omega}, $\Pi_0(\omega_n)$ qualitatively shows the same trend as $\chi^{-1}(\omega_n)$ for small Matsubara frequencies. This suggests that the observed nonlinearity in the SDW susceptibility is likely due to low-energy features of the band structure.

Finally, we show the dependence of the inverse SDW susceptibility on squared momentum in Fig.~\ref{fig:nested_chi_q}. For finite $\mathbf{q} > 0$ the DQMC results for $\chi^{-1}(\mathbf{q}, i\omega_n = 0)$ are consistent with a $q^2$ term, while a noticeable drop is visible at $\mathbf{q} = 0$. We perform a linear regression to establish the quadratic momentum dependence, illustrated in Fig.~\ref{fig:nested_chi_q}, and find good agreement with the numerical data over the range $0<q^2\leq 2$. Importantly, all finite-$q$ data points collapse onto a single line and do not branch out as a function of squared momentum. The isotropic behavior of the boson propagator lies in stark contrast to an anisotropic $|q_x +q_y|+|q_x -q_y|$ term predicted at the $z=1$ fixed point \cite{Schlief2017}. However, the apparant discontinuity at $q=0$ indicates there are features at low momenta that cannot be resolved for the available system sizes. In particular, the $z=1$ anistropic behavior may emerge at smaller values of $q$. 
We note that the momentum dependence is similar at higher temperature ($T=0.025$, see inset to Fig. ~\ref{fig:nested_chi_q}).
\begin{figure}
	\centering
	\includegraphics[width=0.47\textwidth]{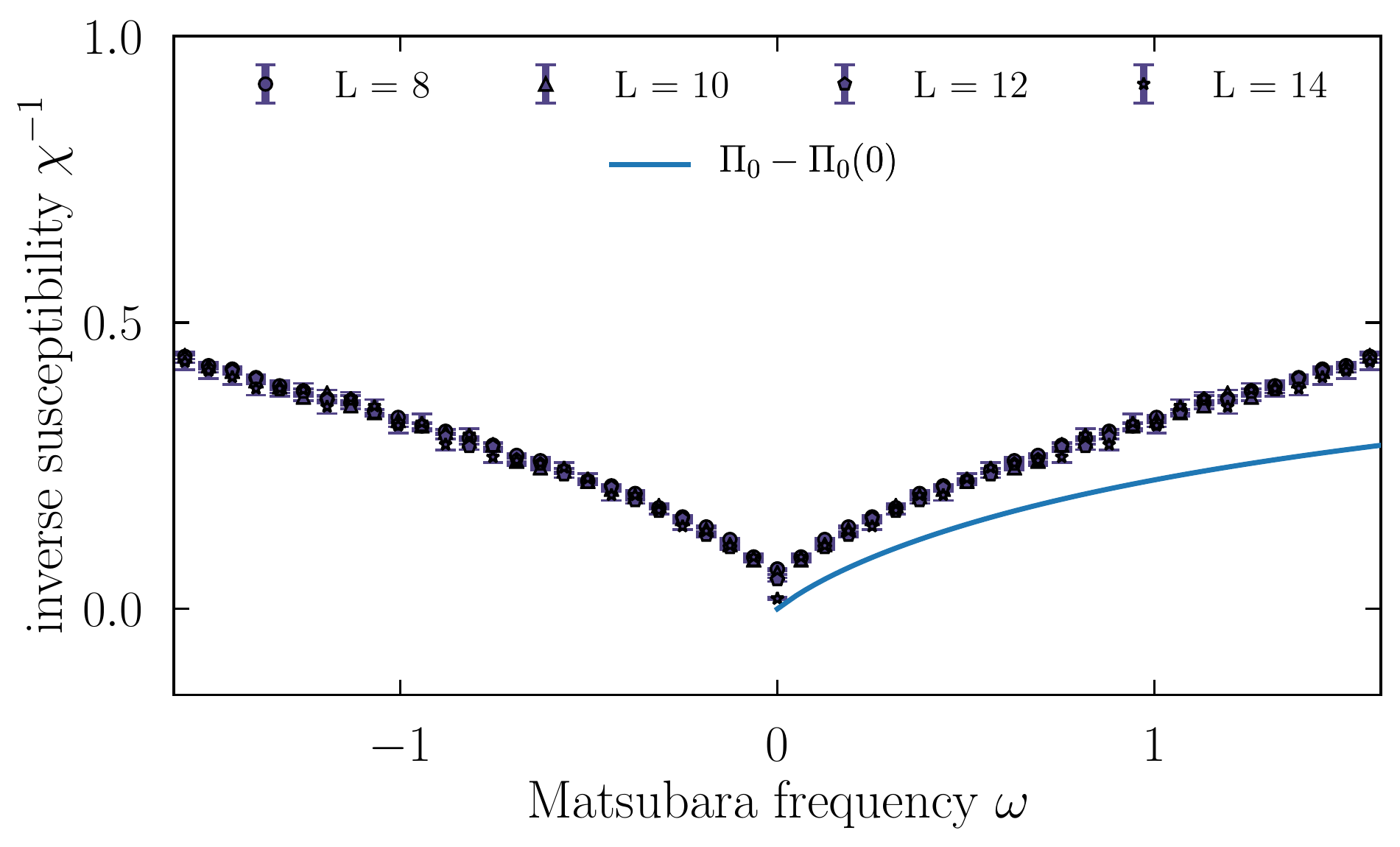}
	\caption{\textbf{Frequency dependence} of the magnetic susceptibility $\chi$ close to the quantum critical point ($r=1.625$) at ultra low temperature $T=1/100$ and $\lambda = 1.0$ for the almost locally nested Fermi surface in Fig. \ref{fig:fermisurface_nested}. The solid line is the non-interacting fermionic susceptibility $\Pi_0$.}
	\label{fig:nested_chi_omega}
\end{figure}

\begin{figure}
	\centering
	\includegraphics[width=0.47\textwidth]{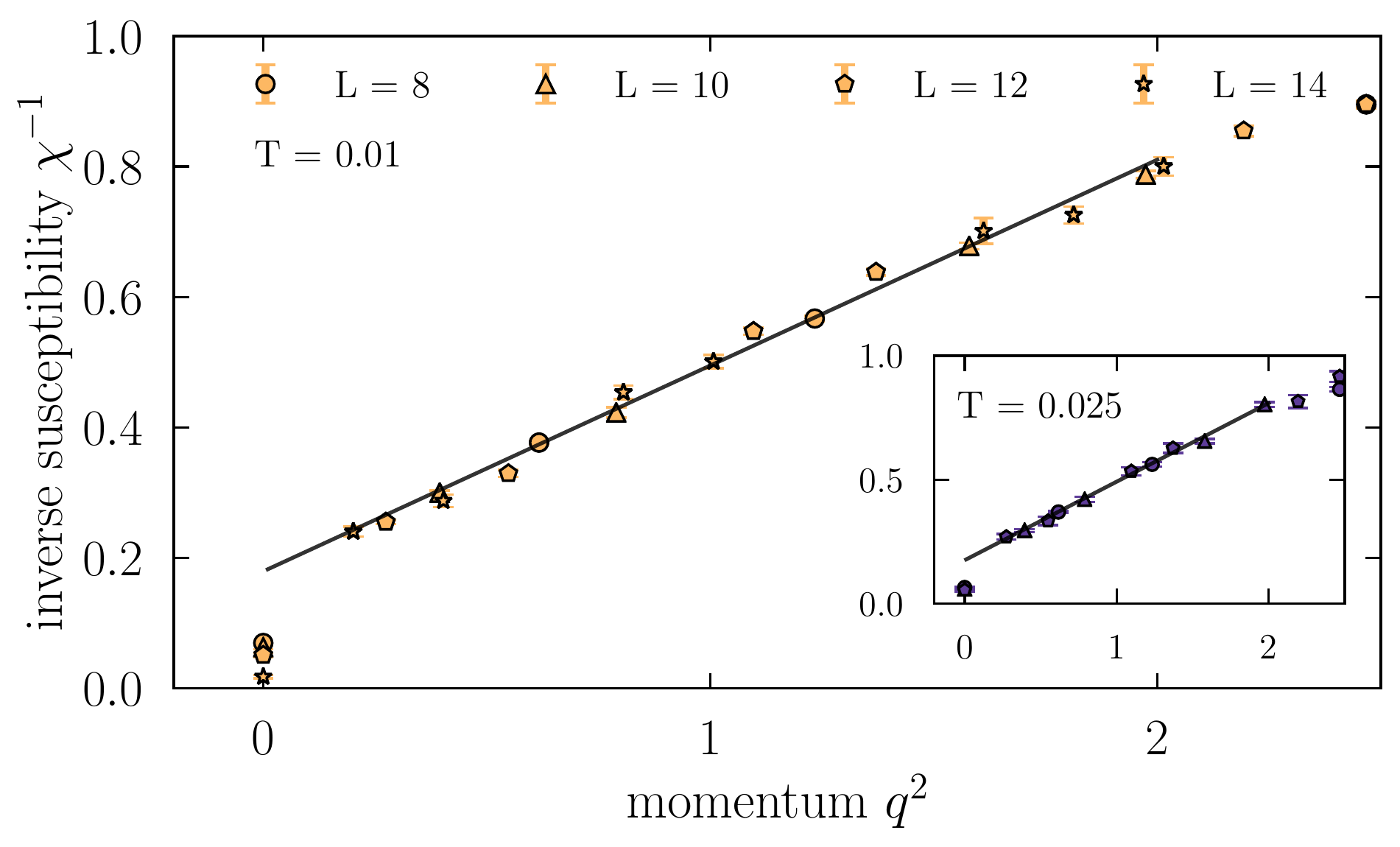}
	\caption{\textbf{Momentum dependence} of the magnetic susceptibility $\chi$ close to the quantum critical point ($r=1.625$) at ultra low temperature $T=1/100$ ($T=1/40$ in the inset) for the almost locally nested Fermi surface in Fig. \ref{fig:fermisurface_nested} with $\lambda = 1.0$. The solid lines are linear fits. 
	}
	\label{fig:nested_chi_q}
\end{figure}

\section{Discussion}
\label{sec:discussion}

In this work, we have deployed large-scale DQMC simulations to explore the phase diagram of itinerant fermions coupled to an isotropic antiferromagnetic order parameter in an unbiased and rigorous manner. By inspecting fermionic and bosonic correlations in the vicinity of the QCP, which marks the onset of SDW order, we were able to unveil the critical low-energy behavior up to numerical accuracy.

Our main finding is that, over a broad range of temperatures and the tuning parameter near the QCP, the critical SDW fluctuations are remarkably well described by Hertz-Millis theory, in which one naively integrates out the fermions~\cite{Hertz1976,Millis1993}. The dynamical exponent in this regime, within our accuracy, is $z=2$.  
This is surprising in view of the fact that Hertz-Millis theory neglects infinitely many marginal couplings that are generated when integrating over the gapless fermion degrees of freedom at the Fermi surface \cite{Abanov2004}. Qualitatively similar observations have been made for metals with easy-plane $XY$ \cite{Gerlach2017} and $Z_2$ antiferromagnetic order \cite{Liu2018}, despite noticeable differences in the perturbative structure \cite{Metlitski2010}, which hints towards a generic property of the SDW QCP. 

A further important aspect of our DQMC results is that, away from perfect local nesting at the hotspots, the QCP always appears to be shadowed by high-temperature superconductivity. We find that the superconducting order parameter is unambiguously of $d$-wave character, i.e. changes sign under $\pi/2$ rotations. The observed maximal critical temperature $T_c^{\mathrm{max}} \approx E_F/20 $ for $\lambda=2$ is in good agreement with Eliashberg theory. Anticipating that the energy scale associated with superconductivity due to fermions at the hotspots, Eq.~\ref{eq:scscale}, is independent of the band structure parameters \cite{Wang2016} our $T_c$ can be compared to Ref.~\cite{Schattner2016,Gerlach2017}, which considers a very similar model with easy-plane AFM fluctuations. We find $T_c^{O(3)} / T_c^{O(2)} \approx 2.2$, consistent with a quadratic dependence of $T_c$ on the number of order parameter components.

In much of the ``quantum critical'' regime above the superconducting $T_c$, the fermions at the hot spots remain underdamped, in the sense that their self-energy is smaller than their energy, except at the lowest temperatures approaching $T_c$.  
Above the QCP, the imaginary part of the self-energy at the hot spots is only weakly dependent on Matsubara frequency, in stark contrast to the behavior expected within Fermi liquid theory. On the ordered side of the QCP the self-energy seems to be diverging, which is consistent with a gapping out of the hot spots.

Motivated by the recent prediction of a novel $z=1$ fixed point \cite{Schlief2017, Lunts2017, Lee2018} by 
Sung-Sik Lee and collaborators, we considered a variant of the model where the hot spots are (nearly) locally nested. In this case, the superconducting $T_c$ is strongly suppressed below the lowest temperature available in our study, $T \approx E_F/200$. Above this temperature, we find substantial deviations from the $z=2$ Hertz-Millis behavior, but no evidence for the predicted $z=1$ criticality. Notably, the momentum dependence of the SDW order parameter is isotropic up to numerical accuracy.

A further investigation of the possibility of a $z=1$ fixed point manifestation at even lower temperatures and with different band structures is desirable. Moreover, directly extracting the renormalized vertex function from DQMC simulations and analyzing its Matsubara frequency dependence seems like a viable approach to shed important light onto why Hertz-Millis theory is capable of capturing large parts of the critical physics correctly, despite being formally uncontrolled. On a different front, the recent advances in the field of quantum machine learning \cite{Carrasquilla2017,Broecker2017,Zhang2017,VanNieuwenburg2017,Carleo2016} provide a new exciting route to accessing transport properties \cite{Zhang2019} and might provide a computationally efficient way to identify and map out an extended non-Fermi liquid regime.

\begin{acknowledgements}
We acknowledge useful discussions with Steven Kivelson, Sung-Sik Lee, Peter Lunts, and Xiaoyu Wang.
We acknowledge partial support from the Deutsche Forschungsgemeinschaft (DFG, German Research Foundation) Projektnummer 277101999 -- TRR 183 (projects A01, B01). 
The numerical simulations were performed on the JUWELS cluster at FZ Juelich, the CHEOPS cluster at RRZK Cologne and on the Sherlock cluster at Stanford. EB was supported by the European Research Council (ERC) under grant HQMAT (grant no. 817799), and by the US-Israel Binational Science Foundation (BSF). YS was supported by the Department of Energy, Office of Basic Energy Sciences, under Contract No. DE-AC02-76SF00515 at Stanford, and by the Zuckerman STEM Leadership Program.
This work was supported by a research grant from Irving and Cherna Moskowitz.

\end{acknowledgements}

\bibliography{references}

\appendix

\section{First order transition for $u=0$}
\label{app:firstorder}

It is always a possibility that the onset of magnetism is first order in character rather than continuous. If one turns off the quartic boson couplings in model \eqref{eq:model} one indeed observe a {\it coexistence} of {phases}. In this case, as shown in Fig.~\ref{fig:chi_r_u_zero}, the inverse SDW susceptibility $\chi^{-1}(r)$ is still continuous at temperatures $T \ge 0.1$ but very visibly becomes discontinuous at lower temperature. The fact that this jump-like tuning parameter dependence indeed stems from coexisting phases of similar energy can be established by inspecting $\chi$-histograms which clearly reveal a double-peak structure (not shown).

\begin{figure}[t]
	\centering
	\includegraphics[width=0.47\textwidth]{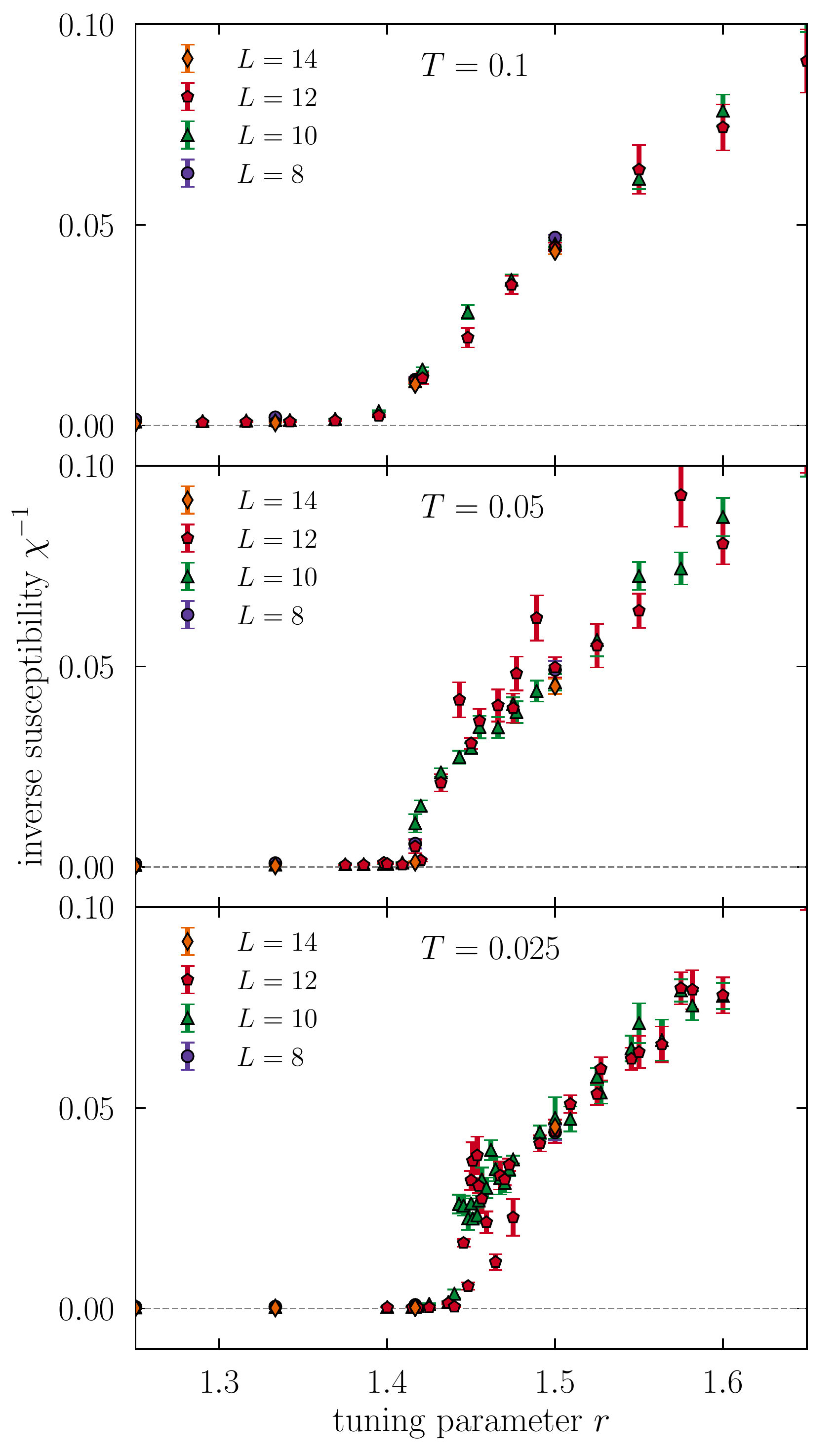}
	\caption{\textbf{First order transition} occurring at low temperatures for $u=0$. Shown is the tuning parameter dependence of the inverse magnetic susceptibility $\chi^{-1}$ close to the quantum critical point at a couple of temperatures.}
	\label{fig:chi_r_u_zero}
\end{figure}

Being interested in quantum critical properties we stayed away from this first-order transition in the work presented in the main text. Note that the almost locally nested system of Sec.~\ref{sec:nesting} has $u=0$ and might be argued to exhibit a first-order transition on general grounds \cite{Altshuler1995}, but nonetheless our numerics shows a continuous onset of SDW correlations down to temperatures as low as $T=0.025$, as evident in Fig.~\ref{fig:nested_chi_r}.

\begin{figure}
	\centering
	\includegraphics[width=0.47\textwidth]{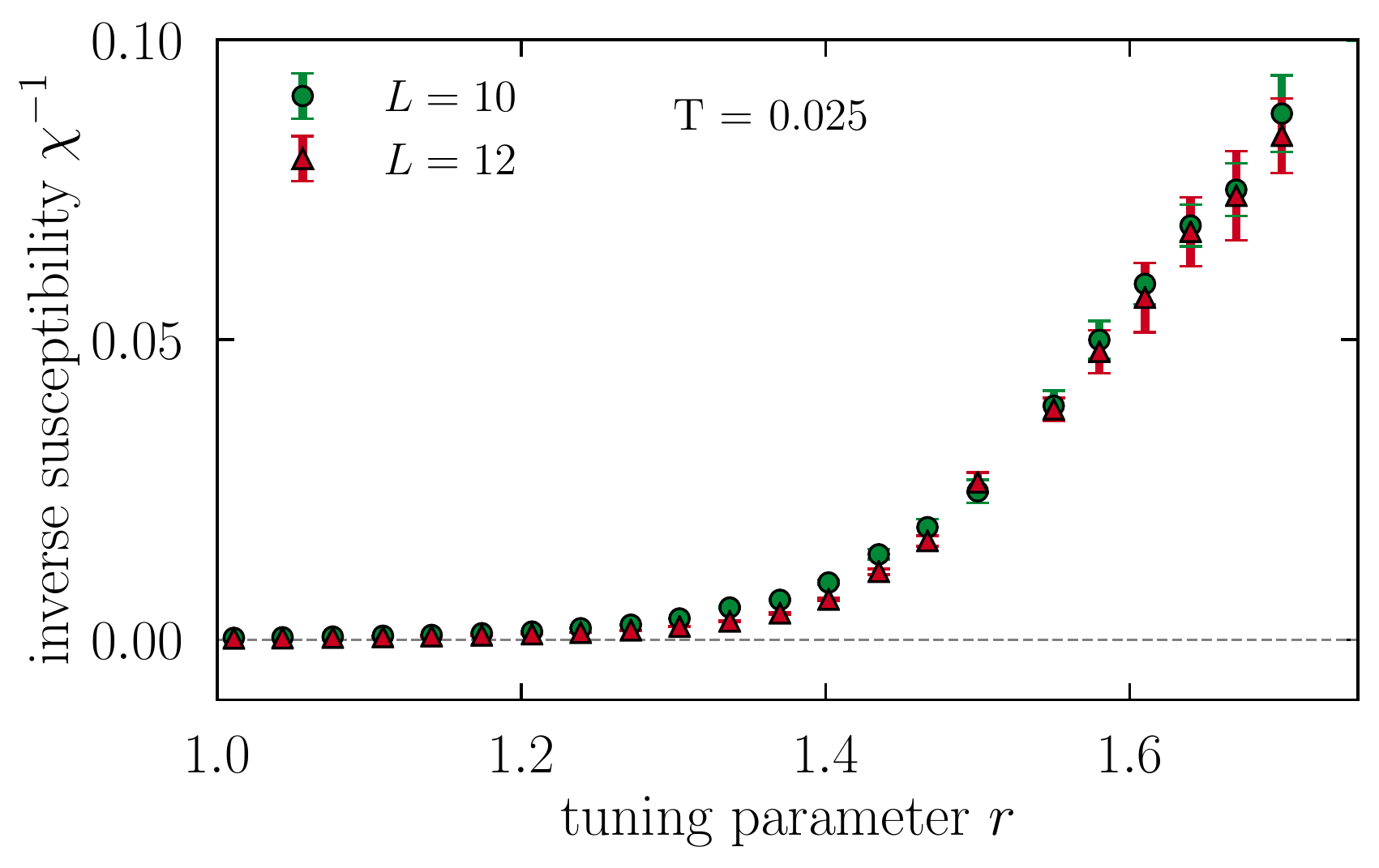}
	\caption{\textbf{Continuous tuning parameter dependence} of model \eqref{eq:model} for the locally nested Fermi surface in Fig.~\ref{fig:fermisurface_nested} with vanishing quartic coupling, $u=0$, at inverse temperature $\beta = 40$.}
	\label{fig:nested_chi_r}
\end{figure}

\section{Applied magnetic field and twisted boundary conditions \label{sec:flux}}

To reduce finite size effects in our simulations we thread a single magnetic flux quantum $\phi_0$ through our system by introducing Peierls phases into the fermion kinetic energy \cite{Assaad2002, Schattner2016, Gerlach2017},
\begin{align}
\mathcal{H} = -t \sum_{\langle i,j \rangle} e^{iA^{\alpha s}_{ij}} c_i^\dagger c_j + \textrm{h.c.}
\end{align}
Here, $A^{\alpha s}_{\mathbf{i}\mathbf{j}} = \frac{2\pi}{\phi_0} \int_\mathbf{i}^\mathbf{j} \mathbf{A}^{\alpha s} \cdot \mathbf{dl}$ and we choose a vector potential in Landau gauge, ${\mathbf{A}^{\alpha s}(\mathbf{r}) = -B_{\alpha s}y\mathbf{\hat{x}}}$. To retain the antiunitary symmetry, which renders our model sign-problem free, we choose flavor and spin dependent amplitudes,

\begin{align}
B_{x\uparrow} = B_{y\downarrow} = \frac{\phi_0}{L^2} = - B_{x\downarrow} = - B_{y\uparrow}.
\end{align}

Note that this choice explicitly breaks the SU(2) spin rotation symmetry of the model. However, the symmetry breaking field scales as $L^{-2}$, rendering it unimportant as long as the antiferromagnetic correlation length is $\xi \lesssim L$. Numerically, we find that the parallel and perpendicular components of the magnetic susceptibility agree (up to error bars) for $r>r_{\mathrm{QCP}}$ and only start to diverge at low temperatures on the ordered side of the QCP.

Explicitly, we implement the following phases for nearest, next-nearest, and next-next nearest neighbor hoppings (Fig.~\ref{fig:square_lattice} in the main text).

\textit{Nearest neighbor}
\begin{align}
A^{\alpha s}_{\mathbf{i}\mathbf{j}} = \begin{cases} 
-\frac{2\pi}{\phi_0} _{\alpha s}B i_y, & \leftarrow \text{hopping}, \\
+\frac{2\pi}{\phi_0} B_{\alpha s} i_y, & \rightarrow \text{hopping}, \\
0, & \uparrow, \downarrow \text{hopping}, \\
-\frac{2\pi}{\phi_0} B_{\alpha s} L i_x, & \uparrow \text{boundary hopping}, \\
+\frac{2\pi}{\phi_0} B_{\alpha s} L i_x, & \downarrow \text{boundary hopping}. \\
\end{cases} \label{eq:nn}
\end{align}

\textit{Next-nearest neighbor}
\begin{align}
A^{\alpha s}_{\mathbf{i}\mathbf{j}} = \begin{cases}
-\frac{2\pi}{\phi_0} B_{\alpha s} \left(i_y + \frac{1}{2}\right), & \downleftarrow \text{hopping}, \\
+\frac{2\pi}{\phi_0} B_{\alpha s} \left(i_y + \frac{1}{2}\right), & \uprightarrow \text{hopping}, \\
-\frac{2\pi}{\phi_0} B_{\alpha s} \left(i_y + \frac{1}{2}\right), & \upleftarrow \text{hopping}, \\
+\frac{2\pi}{\phi_0} B_{\alpha s} \left(i_y + \frac{1}{2}\right), & \downrightarrow \text{hopping} \\
+\frac{2\pi}{\phi_0} B_{\alpha s} \left(L i_x + \frac{1}{2}\right), & \downleftarrow \text{hopping (boundary crossing)}, \\
-\frac{2\pi}{\phi_0} B_{\alpha s} \left(L i_x + \frac{1}{2}\right), & \uprightarrow \text{hopping (boundary crossing)}, \\
+\frac{2\pi}{\phi_0} B_{\alpha s} \left(L i_x - \frac{1}{2}\right), & \downrightarrow \text{hopping (boundary crossing)}, \\
-\frac{2\pi}{\phi_0} B_{\alpha s} \left(L i_x - \frac{1}{2}\right), & \upleftarrow \text{hopping (boundary crossing)}. \end{cases}
\end{align}

\textit{Next-next-nearest neighbor}
\begin{align}
A^{\alpha s}_{\mathbf{i}\mathbf{j}} = \begin{cases}
-\frac{4\pi}{\phi_0} B_{\alpha s} i_y, & \dashedleftarrow \text{hopping}, \\
+\frac{4\pi}{\phi_0} B_{\alpha s} i_y, & \dashedrightarrow \text{hopping}, \\
0, & \dasheduparrow, \dasheddownarrow \text{hopping}, \\
-\frac{2\pi}{\phi_0} B_{\alpha s} L i_x, & \dasheduparrow \text{hopping (boundary crossing)}, \\
+\frac{2\pi}{\phi_0} B_{\alpha s} L i_x, & \dasheddownarrow \text{hopping (boundary crossing)}.
\end{cases} \label{eq:NN}
\end{align}

\begin{figure}
	\centering
	\includegraphics[width=0.49\textwidth]{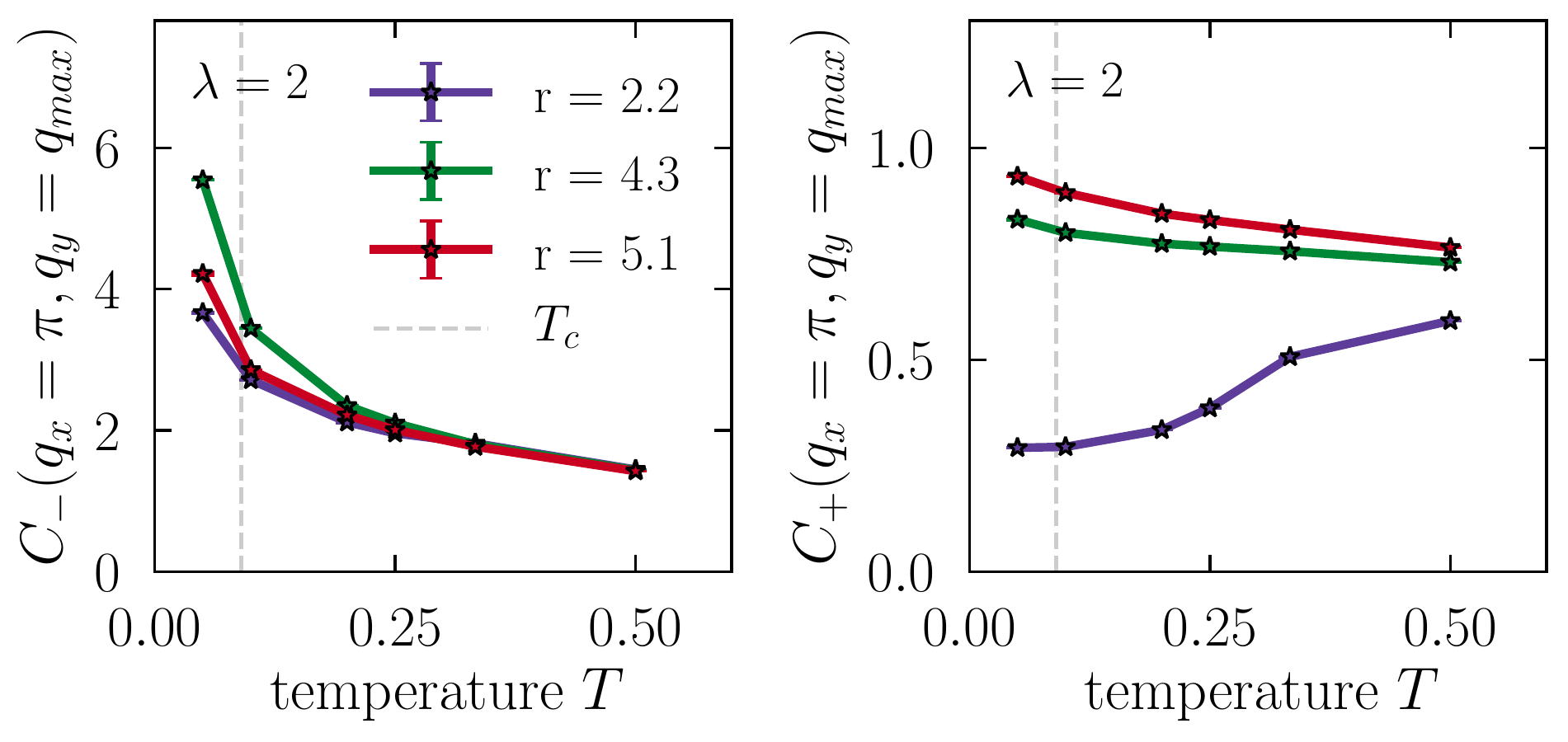}
	\includegraphics[width=0.49\textwidth]{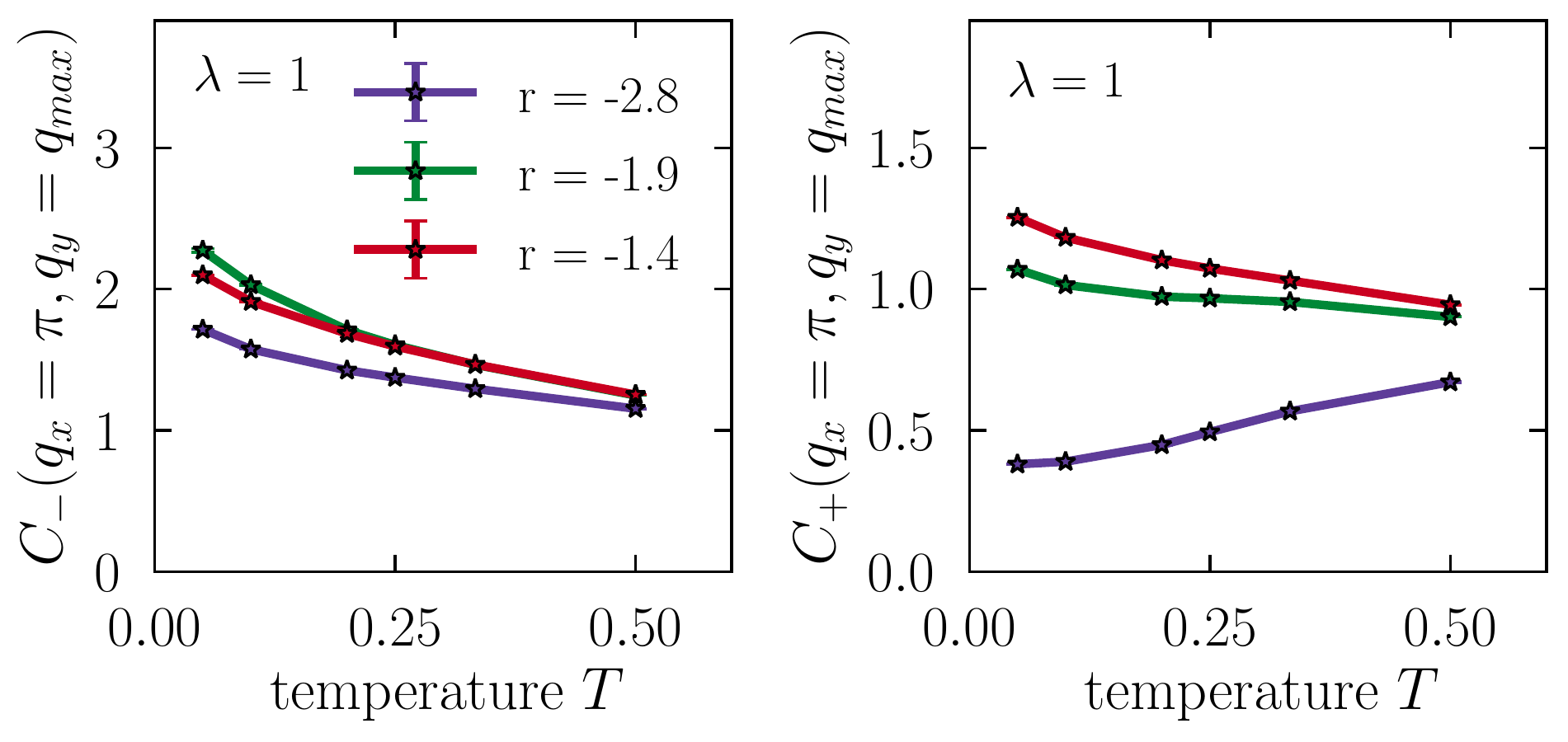}
	\caption{\textbf{Charge-density correlations} across the phase diagram for $\lambda = 2$ (top panel) and $\lambda = 1$ (bottom panel). Shown is the temperature dependence of the maximum (excluding $\mathbf{q}=0$) of the correlations defined in Eq.~\eqref{eq:cdc} for different tuning parameter values $r\ll r_{c}$ (purple),  $r \gtrsim r_{c}$ (green), and $r \gg r_{c}$ (pink). For $\lambda=2$, the dashed, grey line indicates the temperature at which superconductivity sets in (for $r = 4.3 \gtrsim r_{c}$).}
	\label{fig:cdc}
\end{figure}

To study the fermionic Green's function in momentum space, we cannot apply such a flux, but rather use twisted boundary conditions. Here, too, to avoid the sign problem, the twist angles $\varphi$ have an orbital and spin dependence
\begin{align}
\varphi_{x\uparrow} = \varphi_{y\downarrow} = - \varphi_{x\downarrow} = - \varphi_{y\uparrow}.
\end{align}
We use $\varphi=2\pi n /4$ for $n=1...4$ to increase the set of available momenta 16 fold.
To study the self-energy on the Fermi surface, as in Fig. \ref{fig:self_energy} we combine the data from the different boundary conditions, as mentioned above, and identify the local maxima of the $G_\mathbf{k} (\tau=\beta/2)$ at $T=0.2$ as the Fermi surface. The hot spots are the closest point to the intersection of the Fermi surfaces of the two bands $\psi_x$ and $\psi_y$. We take the self energy to be the average of the self-energies at the hotspot and the 4 adjacent momenta (which correspond to different boundary conditions). The error is taken to be the standard deviation across the aforementioned momenta. We follow a similar procedure for $\Sigma{\mathbf{k_x}}$.

\section{Charge-density correlations \label{sec:cdc}}
In order to investigate the possibility of the presence of charge-density wave (CDW) fluctuations we inspect the susceptibilities
\begin{align}
C_{\pm}(r) = \int d\tau \langle \tilde{\Delta}_{\pm}^\dagger(r_i, \tau) \tilde{\Delta}_{\pm}(0,0) \rangle, \label{eq:cdc}
\end{align}
where
\begin{align}
\tilde{\Delta}_\eta = \sum_{\sigma = \uparrow, \downarrow} \psi^\dagger_{xi\sigma} \psi_{xi\sigma} + \eta \psi^\dagger_{yi\sigma} \psi_{yi\sigma}.
\end{align}
In Fig.~\ref{fig:cdc} we show the temperature dependence of the maximum of $C_\pm(\mathbf{q})$, omitting $\mathbf{q}=0$, for three different tuning parameter cuts through the phase diagrams shown in Fig.~\ref{fig:phasediagram}. Both $C_+$ and $C_-$ are peaked close to (but not quite at) the corners of the Brillouin zone at momentum ${\mathbf{q} = (\pi, q_{\textrm{max}} \approx 0.83 \pi})$. First, we notice that $C_+ < C_-$ consistently for all chosen parameters. Second, we observe that while $C_+$ is suppressed with decreasing temperature for $r \ll r_{c}$, indicating a competition with SDW fluctuations, $C_-$ is enhanced for all tuning parameter values. Focusing on $C_-(T)$, we observe that the amplification is most pronounced close to the QCP, $r \gtrsim r_{c}$, and seems insensitive to the onset of superconductivity at $T=T_c$. This behavior is different from that of an easy-plane antiferromagnetic QCP, where the $C_-$ was found to drop sharply below $T_c$~\cite{Schattner2016}. Despite this mild enhancement of CDW correlations at low temperatures in the $O(3)$, we find that their system size dependence across the phase diagram is very weak (not shown), indicating that there is no CDW long-range order.

\section{Superfluid density \label{sec:sfdensity}}

\begin{figure}
	\centering
	\includegraphics[width=0.47\textwidth]{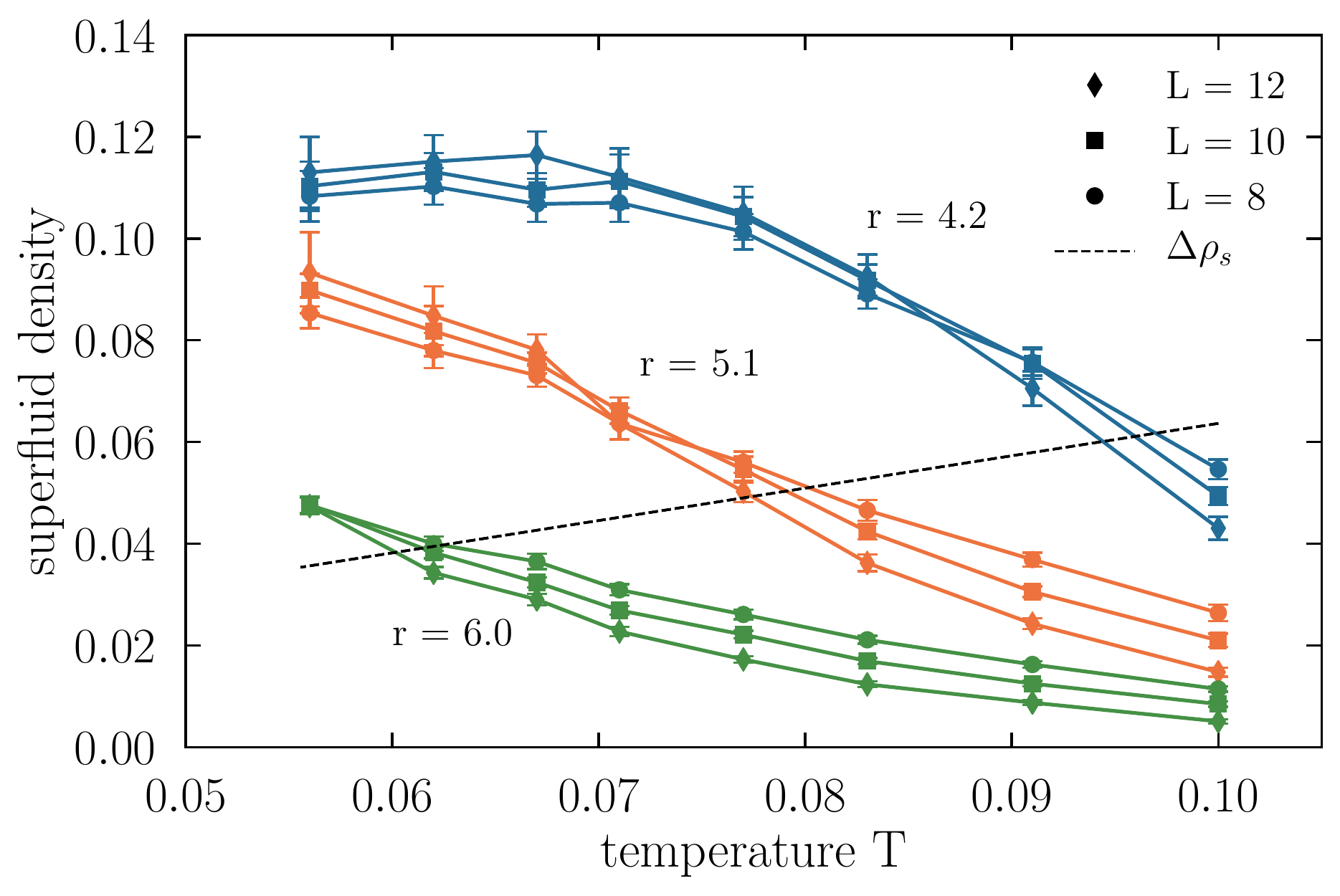}
	\caption{\textbf{Temperature dependence of the superfluid density} of model \eqref{eq:model} for $\lambda=2$. The colors indicate three different tuning parameter values $r = 4.2 \approx r_{T_c^{max}}$ (blue), $r = 5.1 > r_{T_c^{max}}$ (orange), and $r = 6.0 \gg r_{T_c^{max}}$ (green). The dashed line indicates the critical BKT value $\Delta\rho_s = 2T/\pi$.}
	\label{fig:sfdensity}
\end{figure}

To study the onset of superconductivity we apply the criterium by \citet{Scalapino1993}. In particular, we compute the superfluid density of the system from imaginary time current-current correlations, readily accessible in our quantum Monte Carlo simulations, and associate the superconducting transition with those points in the phase diagram where the superfluid density surpasses the BKT value $\Delta\rho_s = 2T/\pi$ (see Ref.~\cite{Schattner2016} for details). In Fig.~\ref{fig:sfdensity} we show the temperature dependence of the superfluid density for three values of the tuning parameter $r$. The biggest source of error is the limitation to finite-size systems. However, based on the available data we anticipate that deviations originating from finite system sizes are typically $\lesssim 10\%$ of the determined transition temperature.

\clearpage

\end{document}